\documentclass[10pt,superscriptaddress,twocolumn,pra]{revtex4-2}
\usepackage{amsthm,amsmath,amssymb}
\usepackage{mathrsfs}
\usepackage{appendix}
\usepackage{amstext}
\usepackage{graphicx}
\usepackage{url}
\usepackage{float}
\usepackage{bm}
\usepackage{ifthen}
\usepackage[usenames,dvipsnames]{color}
\usepackage{mathrsfs}
\usepackage[colorlinks=true, citecolor=blue, urlcolor=black]{hyperref}
\usepackage{dcolumn}
\usepackage{natbib}
\usepackage{booktabs}
\usepackage{hyperref}
\usepackage{color}
\usepackage{multirow}
\usepackage{mathtools}
\usepackage{lipsum}
\usepackage{makecell}
\usepackage{slantsc}

\newcommand{\bra}[1]{\mbox{$\left \langle #1 \right|$}}
\newcommand{\ket}[1]{\mbox{$\left| #1 \right\rangle$}}
\newcommand{\etal}{\mbox{$et$ $al$. }}

\begin{document}
	
\title{Asynchronous quantum repeater using multiple quantum memory}
\author{Chen-Long Li}
\affiliation{National Laboratory of Solid State Microstructures and School of Physics, Collaborative Innovation Center of Advanced Microstrucstures, Nanjing University, Nanjing 210093, China}
\affiliation{Department of Physics and Beijing Key Laboratory of Opto-electronic Functional Materials and Micro-nano Devices, Key Laboratory of Quantum State Construction and Manipulation (Ministry of Education), Renmin University of China, Beijing 100872, China}
\author{Hua-Lei Yin}
\email{hlyin@ruc.edu.cn}
\affiliation{Department of Physics and Beijing Key Laboratory of Opto-electronic Functional Materials and Micro-nano Devices, Key Laboratory of Quantum State Construction and Manipulation (Ministry of Education), Renmin University of China, Beijing 100872, China}
\affiliation{National Laboratory of Solid State Microstructures and School of Physics, Collaborative Innovation Center of Advanced Microstrucstures, Nanjing University, Nanjing 210093, China}
\affiliation{Beijing Academy of Quantum Information Sciences, Beijing 100193, People’s Republic of China}
\author{Zeng-Bing Chen}\email{zbchen@nju.edu.cn}
\affiliation{National Laboratory of Solid State Microstructures and School of Physics, Collaborative Innovation Center of Advanced Microstrucstures, Nanjing University, Nanjing 210093, China}

%\noindent ${\ast}$ These authors contributed equally to this work. \\
\date{\today} 
\begin{abstract}
A full-fledged quantum network relies on the formation of entangled links between remote location with the help of quantum repeaters. The famous Duan-Lukin-Cirac-Zoller quantum repeater protocol is based on long distance single-photon interference, which not only requires high phase stability but also cannot generate maximally entangled state. Here, we propose a quantum repeater protocol using the idea of post-matching, which retains the same efficiency as the single-photon interference protocol, reduces the phase-stability requirement and can generate maximally entangled state in principle. We also outline an implementation of our scheme based on the Kerr nonlinear resonator. Numerical simulations show that our protocol has its superiority by comparing with existing protocols under a generic noise model and show the feasibility of building a large-scale quantum communication network with our scheme. We believe our work represents a crucial step towards the construction of a fully-connected quantum network.
\end{abstract}
	
\maketitle

\section{Introduction}
Building a connected network with quantum resources holds promise in both secure communication~\cite{bennett1992communications,xie2022breaking,jessica2022quantum} and information processing~\cite{arute2019quantum,zhong2020quantum,zhou2022experimental,caccoapuoti2020quantum}, where establishing entanglements over long distances is indispensable. 
Unfortunately, photons, transferred through the quantum channel, attenuate due to the channel loss, imposing limitations on the communication efficiency.
Apart from the photon loss, physical imperfections can induce errors and diminish efficiency as well.
Quantum repeaters are introduced to counteract the loss and operation errors stemming from the optical attenuation and physical imperfections and establish entangled qubits between remote locations~\cite{duan2001long,zhao2007robust,sanguoard2011quantumrepeater,Azuma2023Quantum,munro2010quantum,munro2012quantum,fowler2010surface,muralidharan2014ultrafast,azuma2015all,li2023alphotonic}.

When encoding quantum information into light, multiple degrees of freedom can be exploited.
A qubit can be encoded into the Hilbert subspace spanned by the vacuum state $\ket{0}$ and the single-photon state $\ket{1}$, where the single-photon interference (SPI) can be employed to swap entanglement.
The pioneering work of Duan-Lukin-Cirac-Zoller (DLCZ) protocol introduces a quantum repeater protocol that leverages atomic ensembles as quantum memories~\cite{duan2001long}.
In the DLCZ protocol, an atomic ensemble is illuminated by a laser pulse, leading to the emission of a single Stokes photon in the forward direction. 
The Stokes photons generated from two nodes to be entangled are combined at a beam splitter, where the combined radiation is measured by photodetectors.
A successful click projects the joint system of two atomic ensembles into an almost maximally entangled state.
Given that in the SPI process only one single photon arrives at the beam splitter, the entanglement efficiency of the DLCZ protocol is proportional to $\exp[-L/(2L_{\text{att}})]$ with $L$ the distance between two nodes and $L_{\text{att}}$ the attenuation distance.
Albeit the DLCZ is efficient in entangling, it has inherent drawbacks.
The SPI is sensitive to the phase instability, where the path length phase instabilities should be maintained within a fraction of the photon's wavelength~\cite{zhao2007robust}.
Consequently, the phase locking is required for compensating the phase fluctuation~\cite{yu2020entanglement}, which poses a significant technical challenge.
Furthermore, the SPI in the DLCZ protocol cannot generate a perfect maximally entangled state even without noise due to higher excitation, which poses an inherent restriction on the fidelity of generated entanglement in the ideal case.

Alternatively, a qubit can be encoded into degrees of freedom including the time-bin, polarization, or path of photons, where the two-photon interference (TPI) can be used to swap entanglement~\cite{Weinfurter1994experimental,zhao2007robust,jiang2007fast}.
Two atomic ensembles are encoded as one memory qubit at each node~\cite{zhao2007robust} and are excited simultaneously by write laser pulses.
The two generated Stokes photons at the same node are assumed to have an orthogonal polarization state.
Then the Stokes photons from both nodes are directed to a polarization beam splitter and subjected to Bell state measurement to entangle two neighboring memory qubits.
Contrary to the SPI, the two-photon Hong-Ou-Mandel-type interference does not necessitate stable phase correlation, thus facilitating experimental implementation.
Moreover, the high excitation terms can be automatically eliminated in entanglement swapping, enabling the creation of perfect Bell entanglement in the ideal case.
However, since in the process of TPI there are two photons arriving at the middle station, the entanglement efficiency of the quantum repeater protocol based on TPI is proportional to $\exp(-L/L_{\text{att}})$.

In this work, by leveraging the principle of post-matching in quantum cryptography~\cite{Lu2021efficient,xie2022breaking,zeng2022mode,zhou2023experimental,Zhuexp2023}, we propose a quantum repeater protocol for establishing remote entangled states via an asynchronous Bell state measurement.
Our protocol accomplishes entanglement generation by pairing two successful SPI events.
With the proposed architecture, our protocol retains the same entanglement probability as the SPI protocols, which is proportional to $\exp[-L/(2L_{\text{att}})]$.
Furthermore, our protocol is not constrained by the trade-off between the distribution rate and the fidelity with the maximal entanglement.
Additionally, our protocol alleviates the long-distance phase stability requirements inherent in SPI.
Our work can be applied in various physical systems and we propose an implementation of our scheme in Kerr nonlinear resonator.
By amalgamating the advantages of SPI and TPI, we anticipate that our quantum repeater protocol to provide high efficiency and feasibility for a quantum communication link and to be an essential solution for a fully-connected quantum network.
On the other hand, our work connects the two fields--quantum key distribution and quantum repeater--with the idea of single-photon interference and post-matching. We believe this theoretical progress brings new insight about these two fields and benefits the development of the theory and technology in these areas.

\begin{figure}[tbp!]
		\includegraphics[width=8.5cm]{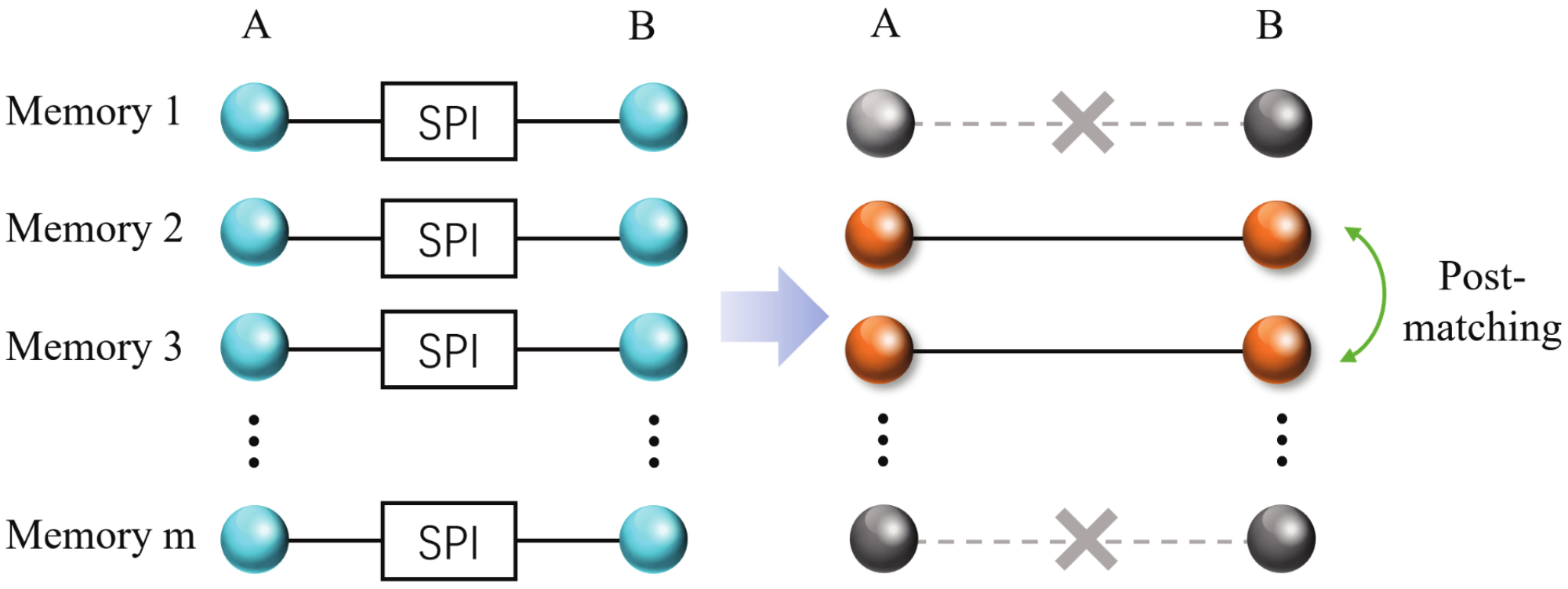}
		\caption{Structure of our protocol in the elementary link. SPI means single-photon interference. At each memory, node A and B prepare entangled states between the qubit (balls shown in the figure) and the optical mode. Two nodes hold the qubits and send the optical modes to the intermediate station to perform SPI. The intermediate station announces the index of a successful SPI events. Two nodes postmatch two events and form a Bell entanglement between them.
		}\label{scheme}
\end{figure}

\section{Quantum repeater protocol}\label{qrep_protocol}
We propose our quantum repeater scheme inspired by the asynchronous post-matching scheme from quantum key distribution~\cite{xie2022breaking,zeng2022mode}.
To be specific, in the asynchronous pairing scheme, Alice and Bob send a train of signals to the intermediate station to perform SPI and then a two-photon Bell state is obtained by post-matching two successful SPI events at different time bins.
In this scheme, the coincidence probability is proportional to $\exp[-L/(2L_{\text{att}})]$ with no requirement on phase locking, which is an impressive feature.

Now we introduce our quantum repeater scheme as shown in Fig.~\ref{scheme}.
At the elementary link, $m$ quantum memories are located at node A which are used to generate qubit-photon entanglement.
At each quantum memory $l$, node A prepares a qubit-photon entangled state $\ket{\phi}^l_{Aa}=\sqrt{p}\ket{e,1}^l_{Aa}+\sqrt{1-p}\ket{g,0}^l_{Aa}$ $(0<p<1)$, where $\ket{e}_{A}$ and $\ket{g}_{A}$ are two eigenstates on the $Z$ basis representing the excited and ground state, respectively.
For a clearer view, in the following illustration of our protocol we use $\ket{\bar{0}}$ and $\ket{\bar{1}}$ to denote $\ket{g}$ and $\ket{e}$.
$\ket{0}_{a}$ and $\ket{1}_{a}$ are the vacuum and single-photon state from node A, respectively.
The above notions and process of node A applies to node B in the same manner.
Node A and B hold their qubits $A$ and $B$ in quantum memories and send the optical modes $a$ and $b$ to perform the SPI in the intermediate station.
The intermediate station announces the position of a successful SPI and the exact result of the SPI.
According to the announcement, node A and B pair two successful SPI events together to generate Bell entanglement between them by performing two-qubit operation and measurements.
For instance, when the SPI are successfully performed between memory $j$ and $k$, the state of the joint system is $\ket{\psi}^{jk}_{AB}=(\ket{\bar{1}\bar{0}}^j_{AB}+e^{i\phi}\ket{\bar{0}\bar{1}}^j_{AB})(\ket{\bar{1}\bar{0}}^k_{AB}+e^{i\phi}\ket{\bar{0}\bar{1}}^k_{AB})/2$, where $\phi$ is the phase difference between the left and the right side channels.
Here memory $j$ and memory $k$ share the same phase difference $\phi$ since all $m$ memories located at each end node generate and send the photons to the intermediate station at the same time.
In the process of post-matching, node A performs {\footnotesize CNOT} gate on the qubit $A$ at the memory $k$ conditioned on the qubit $A$ at the memory $j$.
To be specific, the qubit $A^k$ flips conditioned on the qubit $A^j$ being the state $\ket{\bar{1}}$.
Node B performs the same operation as node A.
The state of the joint system after the {\footnotesize CNOT} gate is $(1/2)(\ket{\bar{1}\bar{0}}^j_{AB}\ket{\bar{0}\bar{0}}^k_{AB}+e^{2i\phi}\ket{\bar{0}\bar{1}}^j_{AB}\ket{\bar{0}\bar{0}}^k_{AB})+(1/2)e^{i\phi}(\ket{\bar{1}\bar{0}}^j_{AB}\ket{\bar{1}\bar{1}}^k_{AB}+\ket{\bar{0}\bar{1}}^j_{AB}\ket{\bar{1}\bar{1}}^k_{AB})$.
Then the qubits $A$ and $B$ at the memory $k$ are measured on the $Z$ basis.
When the measurement result of $A$ and $B$ are both projected on $\ket{\bar{1}}$, node A and B keep the qubits $A$ and $B$ at the memory $j$.
The joint system now is projected onto the Bell state $(1/\sqrt{2})e^{i\phi}(\ket{\bar{1}\bar{0}}^j_{AB}+\ket{\bar{0}\bar{1}}^j_{AB})$ with $\phi$ being the global phase.
Therefore, the long-distance phase stability requirement is relaxed compared with that of SPI in our protocol.
Then high-quality entanglement over longer distance can be established by taking advantage of entanglement swapping and entanglement distillation.
We provide a detailed analysis of state evolution in Appendix~\ref{detailed_ana}.

\begin{table*}[tbp!]
    \setlength{\tabcolsep}{8pt}
	\caption{Comparison among our protocol, protocols based on the single-photon interference (SPI), and protocols based on two-photon interference (TPI). We compare three kinds of protocol in terms of entanglement efficiency, requirements on long-dsitance phase instabilities, whether need global phase locking, and possible physical systems to implement the protocol. In this table, $L_{\text{coh}}$ is a photon's coherence length, which is about 3 m for photons generated from atomic ensembles. $\lambda$ is the wavelength which is about 1 $\mu$m for photons generated from atomic ensembles.}
	%	\centering
	\begin{tabular}{cccc}
		\hline
		\hline
		Scheme& Our protocol & SPI & TPI\\
		\hline
		Entanglement efficiency&$O(\exp[-L/(2L_{\text{att}})])$&$O(\exp[-L/(2L_{\text{att}})])$& $O(\exp(-L/L_{\text{att}}))$\\
        Requirement on phase instabilities&$\le L_{\text{coh}}/10$&$\le \lambda/10$~\cite{zhao2007robust}&$\le L_{\text{coh}}/10$~\cite{zhao2007robust}\\
		Avoid global phase locking&$\surd$&$\times$& $\surd$\\
		Possible physical systems& \makecell[c]{Single atom, atomic \\ensemble, trapped ion,\\ NV center, quantum\\ dot, $etc.$}& \makecell[c]{Single atom~\cite{chou2007functional}, atomic\\ ensemble~\cite{chou2005measurement}, trapped \\ion~\cite{slodic2013atom}, NV center~\cite{humphreys2018deterministic},\\ quantum dot~\cite{delteil2016generation}, $etc.$}&\makecell[c]{Single atom~\cite{julian2012heralded,nolleke2013efficient}, atomic\\ ensemble~\cite{yuan2008experimental}, trapped\\ ion~\cite{stephenson2020high}, NV center~\cite{hensen2015loophole}, $etc.$}\\
		\hline
		\hline
	\end{tabular}	\label{comparison}
\end{table*}

Our protocol retains the same entanglement efficiency as the protocol based on SPI.
If $m$ SPIs are performed at the intermediate station, on average $m\exp[-L/(2L_{\text{att}})]$ successful events can be obtained.
By post-matching two successful SPI events which are phase correlated and establishing a TPI result, we can expect the decay of the pairing rate scales as $(1/2)\exp[-L/(2L_{\text{att}})]$ (see Appendix~\ref{ent_eff_sec} for a detailed derivation of the pairing rate in the asymptotic regime when $m \rightarrow \infty$), which is the same as the scaling of the SPI used in the DLCZ protocol.
On the other hand, another critical feature of our protocol is avoiding global phase locking. The requirement on phase stability makes it necessary to monitor and stabilize the global phase frequently, which complicates the experiment system and reduces the feasibility a lot. Thus, removing global phase locking increases the feasibility and simplifies experiment system.

We summarize the above point in Table~\ref{comparison}, where the requirement on long-distance phase instabilities of our protocol is the same as that of the protocols based on TPI and is looser than that of the protocols based on SPI.
The two fields--quantum key distribution and quantum repeater are connected tightly. The DLCZ quantum repeater protocol~\cite{duan2001long} and the twin-field quantum key distribution~\cite{lucamarini2018overcoming} shares a similar idea of constructing correlated qubits or bits using the SPI. The robust quantum repeater~\cite{zhao2007robust} and measurement-device-independent quantum key distribution~\cite{lo2012mdi} are both based on the TPI. Our scheme uses SPI and post-matching concepts from~\cite{xie2022breaking,zeng2022mode}. The discussion here reveals the connection between the quantum key distribution and the quantum repeater through the photon interference. This new insight will enhance the future progress in both fields.

\section{Implementation}

In the previous section, we only consider the high-level outline of the quantum repeater protocol.
Here in this section, we discuss a possible implementation of our quantum repeater scheme using the Kerr nonlinear resonator--a superconducting resonator off-resonantly coupled to an ancillary superconducting qubit~\cite{leghtas2013hardware}. 
In this system, instead of the states of a superconducting qubit, the quantum information is encoded and protected in a continuous variable system--the superconducting resonator, where we can manipulate the coherent states of the resonator through the interaction between the superconducting qubit and the resonator~\cite{rosenblum2018cnot}. 
This system is superior due to easy experimental implementation, stabilization against loss, and long coherence time~\cite{Kumar2019towards}.

The Hamiltonian of a parametrically driven Kerr nonlinear resonator in a frame rotating at the resonance frequency and in the rotating-wave approximation is given by~\cite{Kumar2019towards, goto2016bifurcation}
\begin{equation}\label{hamcavity}
    H_0=-Ka^{\dagger2}a^2+(\mathcal{E}_pa^{\dagger2}+\mathcal{E}_p^*a^2)
\end{equation}
with $a$ the annihilation operator for the mode of the resonator--the microwave cavity, $K$ the Kerr coefficient, and $\mathcal{E}_p$ the pump amplitude of the parametric drive.
This Hamiltonian can be realized by placing a transmon inside a microwave cavity using suitable microwave tones~\cite{focus2015birgitta,puri2017engineering} or using a $\lambda/4$ transmission line resonator terminated by a flux-pumped SQUID~\cite{puri2017engineering}.
A cat state can be generated via quantum adiabatic evolution.
As the pump amplitude $\mathcal{E}_p$ increases adiabatically, the state of the cavity evolves from the Fock states $\ket{0}$ and $\ket{1}$ to the cat states $\ket{C^+}=\mathcal{N^+}\left(\ket{\sqrt{\mathcal{E}_p/K}}+\ket{-\sqrt{\mathcal{E}_p/K}}\right)$ and $\ket{C^-}=\mathcal{N^-}\left(\ket{\sqrt{\mathcal{E}_p/K}}-\ket{-\sqrt{\mathcal{E}_p/K}}\right)$ respectively, where
$\ket{\sqrt{\mathcal{E}_p/K}}$ and $\ket{-\sqrt{\mathcal{E}_p/K}}$ are coherent states and $\mathcal{N^{\pm}}=1/\sqrt{2\pm2\text{e}^{-2\mathcal{E}_p/K}}$.
We pick $\mathcal{E}_p$ large enough so that the two coherent states have negligible overlap.
By reversing the control pulse, the cat states can be evolved back to the Fock state.
In the following discussion, $\ket{C^+}$ and $\ket{C^-}$ are chosen as logical qubits $\ket{0_l}$ and $\ket{1_l}$ respectively.

The system of Kerr nonlinear resonator can perform universal quantum computation, encompassing both single-qubit and two-qubit gates~\cite{Kumar2019towards}.
In terms of single-qubit gates, the Kerr nonlinear resonator system evolves under
\begin{equation}\label{Xgate}
    H_X=H_0+\mathcal{E}_X(a+a^{\dagger})
\end{equation}
to perform the $X$ rotation with $\mathcal{E}_X$ the amplitude of the photon drive.
The system can be rotated around the $Z$ axis under
\begin{equation}\label{Zgate}
    H_Z=-K(a^{\dagger}a)^2.
\end{equation}
For two-qubit gates, the system evolves under the Hamiltonian for two linearly coupled resonators
\begin{equation}\label{twoqubitgate}
    H_{12}=H_{01}+H_{02}+\mathcal{E}_c(a_1^{\dagger}a_2+a_1a^{\dagger}_2),
\end{equation}
where $H_{0i},(i=1,2)$ is given by Eq.~(\ref{hamcavity}), $a_i$ is the annihilation operator for microwave resonator of system $i$, and $\mathcal{E}_c$ is the coupling strength.
In order to perform a {\footnotesize CNOT} gate, the aforementioned two-qubit gate (denoted as $G_{\theta}$ with $\theta$ the rotation angle) should be combined with the $X$ and $Z$ rotation (denoted as $X_{\theta}$ and $Z_{\theta}$ respectively).
To be specific, the sequence of gates $X^2_{\pi/2}X^1_{-\pi/2}Z^1_{\pi/2}G_{\pi/2}X^1_{-\pi/2}Z^1_{-\pi/2}X^1_{\pi/2}$ applies a {\footnotesize CNOT} gate on qubit 2 conditioned on the state of qubit 1~\cite{Kumar2019towards}.
The $Z$-basis measurement can be implemented by measuring the parity of the cat state by using an ancillary transmon qubit and a fast-decaying readout resonator coupled to the storage resonator~\cite{sun2014tracking}.
Two $\pi/2$ pulses are applied on the transmon, which takes the transmon to either its ground state or excited state according to the parity of the cat state.
Then the parity of the cat state can be read out by reading the state of the transmon.
One can also read the number of photons after mapping the cat state back to the Fock state~\cite{johnson2010quantum}.

We now illustrate how to implement our protocol to generate entanglement between two end nodes at the elementary link.
According to our protocol, there are $m$ multiplexed parts at each node.
We first explain how to generate the qubit-photon entanglement at each part.
At each part, two microwave resonators (we denote as 1 and 2) are coupled together and are initially in the vacuum state $\ket{0}$.
The two resonators start with evolving adiabatically to the cat state $\ket{0_l}$.
Then $X_{\pi/2}$ gate is applied to resonator 1 and the state of the joint system is $(\ket{0_l}_1+\ket{1_l}_1)\ket{0_l}_2/\sqrt{2}$.
{\footnotesize CNOT} is then performed on $\ket{0_l}_2$ conditioned on the state of resonator 1 and the resulting state of the joint system is given by $(\ket{0_l}_1\ket{0_l}_2+\ket{1_l}_1\ket{1_l}_2)/\sqrt{2}$.
By applying the time-reversed control pulse, the cat state of resonator 2 is mapped back to the Fock state and we have the qubit-photon entanglement
\begin{equation}
    \frac{1}{\sqrt{2}}(\ket{0_l}_1\ket{0}_2+\ket{1_l}_1\ket{1}_2).
\end{equation}
The Fock state is then transduced to telecom wavelengths to be coupled into the fiber since the energy of the microwave photons is lower than the thermal noise at room temperature.
After preparing the qubit-photon entanglement at each part, each node sends out photons and the SPI is performed at the intermediate station.
When the intermediate station finishes announcing the result, each end node post-matches two successful SPI events, say $j$ and $k$.
Each end node performs {\footnotesize CNOT} gate on resonator $k$ conditioned on resonator $j$ by evolving the system according to the aforementioned sequence of gates consisting of $X_{\theta}$, $Z_{\theta}$, and $G_{\theta}$.
Then the parity of resonator $k$ is read out as we have described above and the trial is successful when the parity is odd.

Our scheme does not rely on specific transduction protocol.
Here we present a concrete example using rare-earth ion doped crystal proposed in~\cite{Kumar2019towards,electron2019welinski}.
The rare-earth ion doped crystal system has high conversion efficiency due to enhanced coupling strengths and easy integration with fiber.
Specifically, the transduction protocol adopts an Er$^{3+}$ doped Y$_2$SiO$_5$ crystal.
An external constant magnetic field splits the ground state and the excited state.
The optical transitions between ground and excited states are at telecom wavelengths.
The idea of the transduction protocol is first to map the microwave photon to a spin excitation and then map the spin excitation to an optical excitation which is finally read out.
To map the microwave photon to a spin excitation, the collective spin transition couples strongly to a single mode of the microwave cavity, where the Hamiltonian of the system can be described by 
\begin{equation}
    H=\sum_j\hbar\frac{\omega_j}{2}\sigma_z^{(j)}+\sum_j\hbar g'(a^{\dagger}\sigma_-^{(j)}+a\sigma_+^{(j)})+\hbar\omega_ma^{\dagger}a
\end{equation}
with $\sigma_z^{(j)}$ the spin population operator, $\sigma_{\pm}^{(j)}$ the spin flip operator, $\omega_j$ the spin transition frequency of the $j$th spin, $a$ the annihilation operator of the microwave cavity mode, and $\omega_m$ the resonance frequency of the cavity.
The free evolution of the system transfers the cavity excitation into the collective spin excitation.
Then the spin transition is detuned from the cavity and the field gradient is applied.
The emission of a telecom photon is controlled by the dephasing and rephasing of the collective spin and optical dipoles through a $\pi$ pulse sequence.
The transduction efficiency can be greater than 0.85 for realistic parameters~\cite{interfacing2014obrien} and can be further optimized.

According to the entanglement swapping protocol~\cite{gottesman1999demonstrating}, two Bell states with each having one qubit in the middle node, can be deterministically transformed into a Bell state with {\footnotesize CNOT} operation, single-qubit operations, and measurements. In heralded entanglement purification protocol~\cite{deutsch1996quantum}, two Bell states are combined into one Bell state with higher fidelity probabilistically with {\footnotesize CNOT} operation, single-qubit operations, and measurements. Therefore, we can expect that the system can perform entanglement swapping and heralded entanglement purification.

\begin{figure}[tbp!]
		\includegraphics[width=8.5cm]{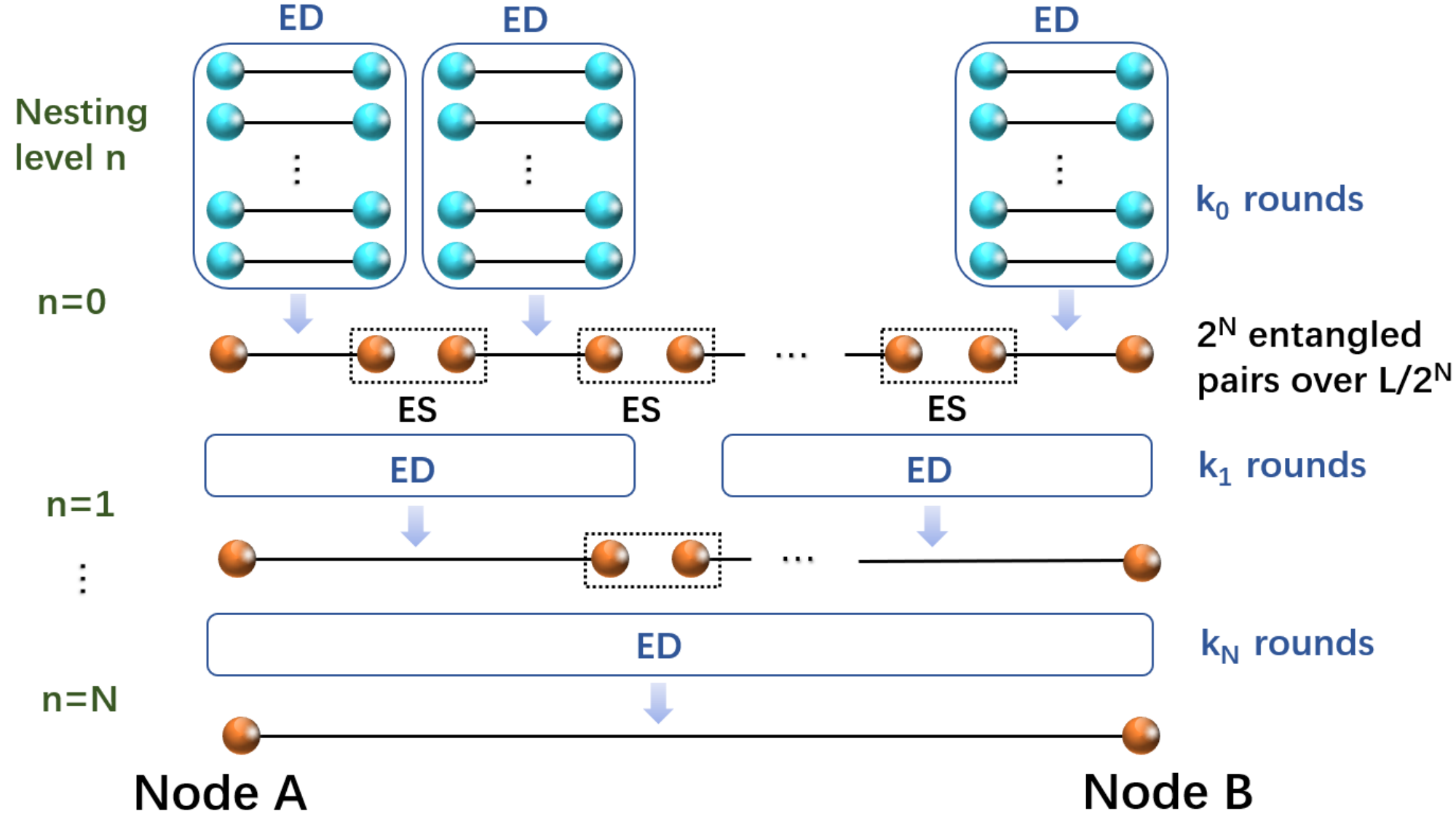}
		\caption{The quantum repeater structure we consider in the numerical simulation. ES and ED are acronyms of entanglement swapping and entanglement distillation, respectively. We can establish the Bell entanglement at $2^N$ elementary links covering $L_0=L/2^N$, perform entanglement swapping at all $N$ nesting levels, and finally obtain the entanglement between node A and B over $L$. Furthermore, entanglement distillation can be involved in our quantum repeater structure as shown in the figure. $k_i$ describes the number of distillation rounds at the $i$-th nesting level.
		}\label{repeater_structure}
\end{figure}
\section{Numerical simulation}

In this section, we analyze the performance of our protocol by numerical simulation.
We first simulate the entanglement generation rate of our protocol under the generic depolarizing noise.
Then we evaluate the fidelity of the generated entanglement based on the Kerr nonlinear resonator system introduced in the previous section under the single photon loss.
We also provide a demonstration of our protocol's advantage in alleviating phase stability requirements.

\subsection{Entanglement generation rate}\label{genration_rate}

We first introduce the quantum repeater structure considered in the simulation.
As shown in Fig.~\ref{repeater_structure}, to establish entanglement between two nodes linked by a quantum channel over long distances, entanglement swapping are used to link the Bell entanglement at the elementary level.
To be specific, an entanglement over the distance $L$ can be generated by the entanglement swapping of two entangled states over $L/2$.
Similarly, we can create the entangled state over $L/2$ by swapping the entanglements over $L/4$.
Therefore, we can establish the Bell entanglement at $2^N$ elementary links covering $L_0=L/2^N$, perform entanglement swapping at $N$ levels, and finally obtain the entanglement between two nodes over $L$.
$N$ is called the nesting level~\cite{sanguoard2011quantumrepeater}.
To improve the fidelity of the generated entanglement, distillation protocols~\cite{dur1999quantum,bennett1996mixed,bennett1996purification,deutsch1996quantum} can be inserted into quantum repeater protocol.
We define an $(N+1)-d$ vector $\Vec{k}=(k_0,..,k_N)$ with each component $k_i$ describing the number of distillation rounds at the nesting level $i$.
In the numerical simulation, we consider the $Oxford$ $protocol$ with local $\pi/2$ $(-\pi/2)$ rotation about the $X$ axis and the {\footnotesize CNOT} operation involved~\cite{deutsch1996quantum}.

To quantify the performance of our proposed quantum repeater protocol, we calculate the repeater rate that is the number of entangled pairs generated between two end nodes over $L$ per second.
The repeater rate can be defined by the reciprocal value of the time $T_0\tau(k_N,N)$ needed to establish entangled pair between two end nodes over long distance.
$T_0=L_0/c$ with $L_0=L/2^N$ and $c$ the speed of light in optical fiber is the time of classical communication over $L_0$.
$N$ is the maximal number of nesting level and $k_N$ is the corresponding number of distillation round.
For the time needed in the elementary link, $\tau(0,0)^{-1}$ is proportional to the success probability of generating a Bell pair, where the fiber loss $\text{exp}(-L/2L_{\text{att}})$ is included.
In Appendix~\ref{avr_time}, we present how to derive the expression of $\tau(k_N,N)$.

In the realistic scenario, the existence of noise will affect the quality of generated entanglement. 
To take the noise in consideration, we adopt a generic gate model with the depolarizing noise~\cite{briegel1998quantum, dur1999quantum}.
For instance, a noisy gate $O_{12}$ acts on a two-qubit state $\rho_{12}$ following $O_{12}(\rho_{12})=p_GO_{12}^{\text{ideal}}(\rho_{12})+(1-p_G)\mathbb{I}_{12}/4$, where $p_G$ is the gate quality.
In our simulation, we consider the effect of an imperfect {\footnotesize CNOT} gate, which is involved in our proposed protocol, the entanglement swapping, and the $Oxford$ distillation protocol.
With the depolarizing model, we can calculate how the entanglement evolves during the quantum repeater protocol and obtains the repeater rate.
See Appendix~\ref{evolution} for details of the evolution of the entangled state.

\begin{figure}[tbp!]
		\includegraphics[width=8.5cm]{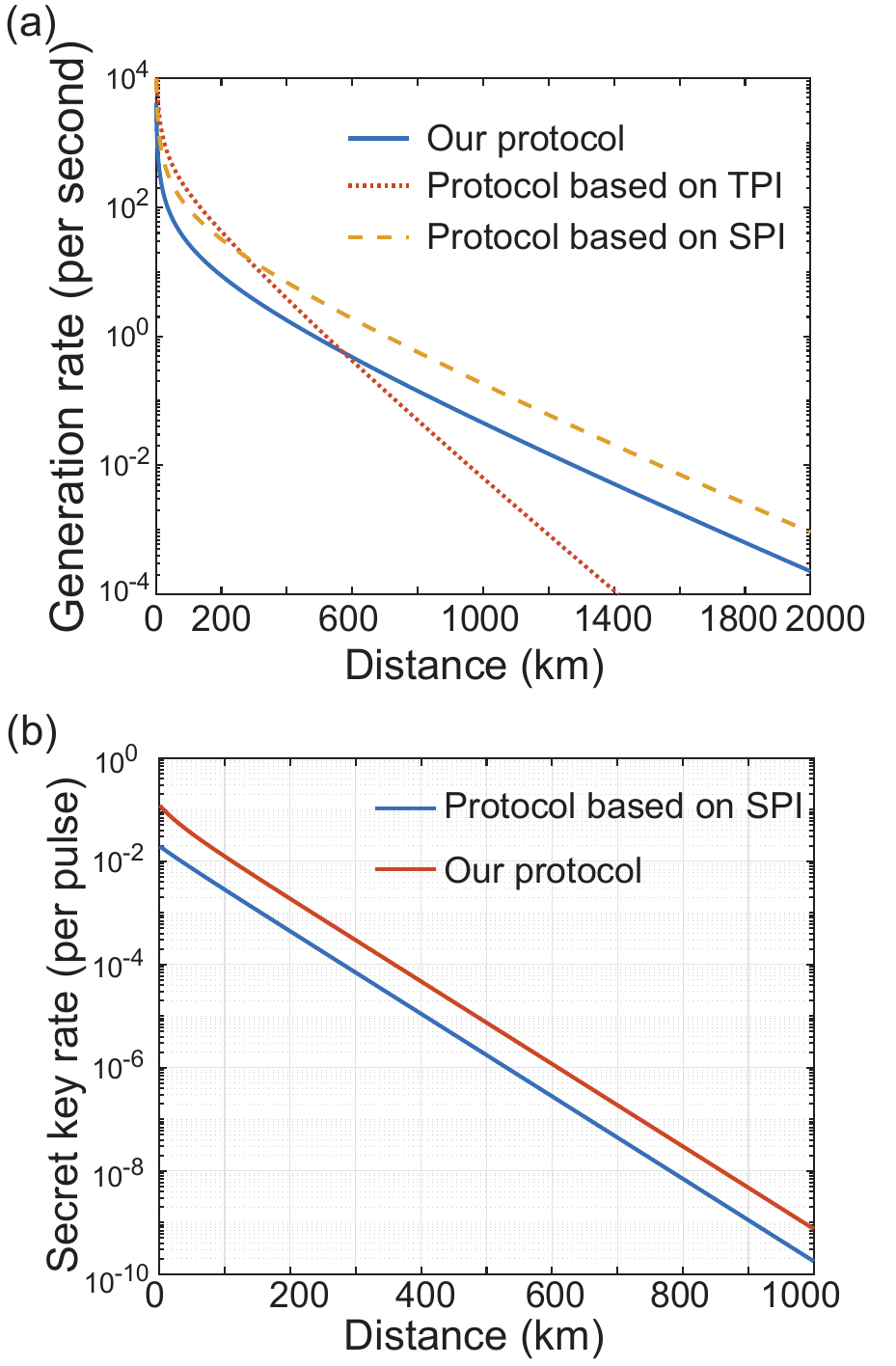}
		\caption{Numerical simulation of entanglement generation rate and secret key rate. (a) The entanglement generation rate at different distances of our protocol, protocol based on SPI~\cite{duan2001long}, and protocol based on TPI~\cite{zhao2007robust} with $N=2$, $\Vec{k}=(2,0,0)$, and $p_G=0.995$. We plot the curve of our protocol with solid line and plot other protocols with dashed lines. (b) Asymptotic quantum key distribution rate of our protocol and single-photon interference protocol. For SPI protocol, we choose $p=0.01$. For our protocol, we choose $p$ that can maximize the asymptotic quantum distribution key rate at every distance. We choose $L_{\text{att}}=27.16$ km in the simulation.
		}\label{comp_rate}
\end{figure}

\begin{figure*}[tbp!]
		\includegraphics[width=17cm]{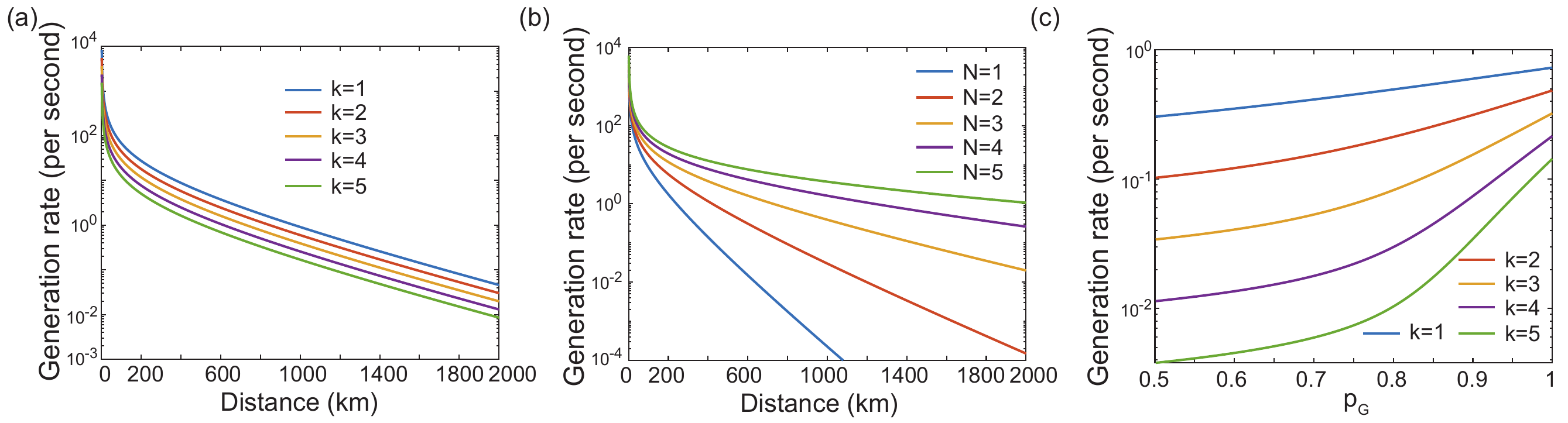}
		\caption{Numerical simulation of our protocol. (a) Rate-distance curves of our protocol under various numbers of distillation round with $N=3$ and $p_G=0.995$. (b) Rate-distance relation of our protocol under different maximal number of nesting level with $k=3$ and $p_G=0.995$. (c) Rate-$p_G$ relation of our protocol under different value of $k$ with $N=2$ and $L=600$ km.
		}\label{num_sim}
\end{figure*}

We present the simulation result under depolarizing noise in Fig.~\ref{comp_rate} (a) and Fig.~\ref{num_sim}.
In the numerical simulation, we consider a simple distillation strategy, where the distillation is performed only before the first-level entanglement swapping, which indicates $\Vec{k}=(k,0,...,0)$.
Thus we use the parameter $k$ to describe the total number of the first distillation round in our numerical result.
As shown in Fig.~\ref{comp_rate} (a), we plot the entanglement generation rate-distance relation of our protocol, protocol based on SPI~\cite{duan2001long}, and protocol based on TPI~\cite{zhao2007robust} with $N=2$, $k=2$, and $p_G=0.995$ (See Appendix~\ref{review_rep} for details of protocols based on SPI and TPI).
From our result, our protocol shares a similar slope with the protocol based on SPI, while the protocol based on TPI shows a lower slope and rate at long distances.
Furthermore, our protocol reduces the requirement of phase stabilization as the protocol based on TPI.
Therefore, this figure shows the most critical feature of our protocol.
The rate of our protocol is lower than the protocol based on SPI since we assume an ideal Bell entanglement at the elementary link for protocols based on SPI and TPI.
We now present the advantage of our protocol from the perspective of quantum key distribution.
Specifically, we calculate the asymptotic secret key rate, where both nodes measure their qubits in $X$ or $Z$ basis to generate secret keys.
The asymptotic secret key rate is $R=Q[1-h(e_X)-h(e_Z)]$~\cite{curty2019simple}, where $Q$ is the total success probability, $e_X$ is the bit error rate, $e_Z$ is the phase error rate, and $h(x)$ is the binary entropy function.
For SPI protocol, we consider both nodes share a noisy entanglement shown in Appendix~\ref{detailed_ana}.
Then we have $Q=2P_{10}$, $e_X=(1-p_1)/2$, and $e_Z=1-p_1$.
$e_X$ is defined by the probability with which node A's and B's $X$ measurement outcomes are different.
$e_Z$ is defined by the probability with which node A's and B's $Z$ measurement outcomes coincide.
For our protocol, since the post-matching operation distills a maximal entanglement, we have $e_X=e_Z=0$.
According to the entanglement efficiency calculated in Appendix~\ref{ent_eff_sec}, we have $Q=p_mp_S/2=P_{10}p_1^2/2$ when pairing two noisy entanglements together.
We present the secret key rate of SPI protocol and our protocol in Fig.~\ref{comp_rate} (b), where we choose $p=0.01$ for SPI protocol and $p$ that maximizes the asymptotic secret key for our protocol.
One can observe our protocol is superior to SPI protocol in quantum key distribution rate.
In SPI protocol, there is a trade-off between the distribution rate and the fidelity with the maximal entanglement.
To be specific, if one wants to suppress errors due to multiple emissions and improve the final fidelity, then one has to work with very low $p$, resulting in a limited distribution rate~\cite{sanguoard2011quantumrepeater}.
In contrast, in our protocol, post-matching operation distills the noisy entanglement, where we obtain a maximal entanglement with a unit fidelity.
Therefore, our protocol is not limited by this trade-off between the distribution rate and the fidelity.
We can improve $p$ to achieve a higher distribution rate without sacrificing the fidelity, which is another critical feature of our protocol.
In Fig.~\ref{num_sim} (a), we fix the maximal nesting level of our protocol $N=3$ and plot the rate-distance relation with various numbers of distillation round $k$.
The entanglement generation rate decreases as $k$ increases since more distillation rounds means longer time consumption.
We plot the rate-distance relation with different nesting levels and fixed $k$ in Fig.~\ref{num_sim} (b).
Because a larger $N$ means a shorter distance of elementary distance $L_0=L/2^N$, the slope of the curve increases as $N$ grows, which means more repeater station can enhance the communication efficiency.
In Fig.~\ref{num_sim} (c) we show the rate-$p_G$ relation.
$p_G$ affects the fidelity of generated entangled pair, which will cause error in cryptographic application~\cite{abruzzo2013quantum,bratzik2013quantum}.

\subsection{Entanglement fidelity}\label{entanglement_fidelity}

In this section, we evaluate the fidelity of generated entanglement based on the system of Kerr nonlinear resonator in the presence of single photons leaking out of the resonator at a rate $\kappa$.
The simulations in this section are carried out using Quantum Toolbox in Python~\cite{johnasson2012qutip}.

Now we analyze and quantify the fidelity of different operations in our scheme.
As the pump amplitude $\mathcal{E}_p$ increases adiabatically, the Fock states $\ket{0}$ and $\ket{1}$ evolves to the cat states $\ket{0_l}$ and $\ket{1_l}$.
In this process, we assume the pump pulse satisfies $\mathcal{E}_p(t)=2K(1-\text{e}^{-(t/\tau)^4})$ to create a cat state with $\sqrt{\mathcal{E}_p/K}=\sqrt{2}$~\cite{Kumar2019towards,puri2017engineering}.
This is a smooth pulse shape and can be optimized to reach the final cat state with a higher fidelity in a faster way~\cite{machnes2011comparing}.
When we adopt $K=2\pi\times0.521$MHz and $\kappa=2\pi\times5.21$Hz~\cite{Kumar2019towards}, we can calculate the fidelity of mapping $\ket{0}$ to $\ket{0_l}$ is 99.87\% in the duration $t=1.3\tau=0.04$ms.
With the single photon drive $\mathcal{E}_X$, $\pi/2$ rotation around $X$ axis can be implemented by evolving the system under Eq.~(\ref{Xgate}) in the duration $t=\pi/(8\sqrt{2}\mathcal{E}_X)$.
In terms of the choice of $\mathcal{E}_X$, there is a tradeoff between faster operation and higher fidelity.
According to~\cite{Kumar2019towards}, we choose $\mathcal{E}_X=2K/45$
and the fidelity of $X_{\pi/2}$ gate is 99.995\%.
$Z_{\pi/2}$ gate can be achieved by evolving the system under Eq.~(\ref{Zgate}) in time $t=\pi/(2K)$ and the fidelity is 99.997\%.
For two-qubit entangling gate $G$, we can evolve the system initially from $\ket{0_l}\otimes\ket{0_l}$ to the maximally entangled state $(1/2)((1+i)\ket{0_l}\otimes\ket{0_l}+(1-i)\ket{1_l}\otimes\ket{1_l})$ in $t=\pi/(16\mathcal{E}_c)$ under Eq.~(\ref{twoqubitgate}).
Similarly, we choose $\mathcal{E}_c=2K/55$ considering the tradeoff between between faster operation and higher fidelity.
The fidelity of the two-qubit entangling gate $G$ is 99.95\%.
To perform {\footnotesize CNOT} operation, additional single-qubit operations are needed~\cite{Kumar2019towards} and the fidelity of {\footnotesize CNOT} operation is 99.7\%.
We take the fidelity for transduction as 99.95\%~\cite{Kumar2019towards}.

The fidelity of the final entanglement between Node A and B can be given by subsequently multiplying the fidelities of different operations and the residual coherence of the storage resonator
\begin{equation}
    F_{\text{AB}}=(\prod_i F_i)^{2^N}\times(F_{\text{ES}})^{2^N-1}\times C_{\text{R}},
\end{equation}
with $\prod_iF_i$ the fidelity of local operations to generate entanglement at the elementary link as we have discussed above, $F_{\text{ES}}$ the fidelity of entanglement swapping, $2^N$ the number of elementary links, and $C_{\text{R}}=\text{e}^{-\kappa_{\text{eff}}T_0\tau(k_N,N)}$ the residual coherence.
$\kappa_{\text{eff}}\approx4\kappa$ is the decoherence rate and $T_0\tau(k_N,N)$ is the average waiting time to distribute entanglement.
In order to increase the residual coherence, error-correction code can be used~\cite{leghtas2013hardware}.
An easier way is to undrive the resonator from cat state to Fock state while waiting, where the decoherence rate is $\kappa$.
Furthermore, the photon from the cat state can be transferred to another high-Q cavity to increase the residual coherence.

We now evaluate the performance of entanglement fidelity of our scheme.
For simplicity, we consider the fidelity of generated entanglement at an elementary link with different distances.
We assume the resonator is undriven to the Fock basis while waiting and thus the decoherence rate is $\kappa$.
In our scheme, 10 driving operations, 4 $X_{\pi/2}$ operations, 6 {\footnotesize CNOT} operations, 6 undriving operations, and 4 transduction operations are needed to generate entanglement at one elementary link.
Therefore, based on the above discussion, we have $\prod_iF_i=96.48\%$.
When $L_0=$10 km, 25 km, and 50 km, we have $F_{AB}=$94.47\%, 88.4\%, and 61.08\%.
We can observe that the entanglement fidelity at one elementary link exceeds the 50\% bound of entanglement purification, where we can expect the fidelity to be further improved in the repeater structure.
Our result shows the feasibility of deploying a quantum network using our scheme in an intra- or inter-city scale.

We provide a brief quantitative demonstration of our protocol’s advantage in alleviating phase stability requirements.
We assume the distribution of the phase fluctuation of the entanglement generated from SPI follows a uniform distribution without phase stabilization~\cite{yu2020entanglement}.
Consequently, the average fidelity of this entangled state with respect to the maximal entanglement is given by $p_1/2$, where $p_1=\frac{p(1-p)\eta_t}{p(1-p)\eta_t+p^2(1-\eta_t)\eta_t}$ with $\eta_t=\text{exp}(-L/(2L_{\text{att}}))$ is attributed to the noise due to the higher excitation term and $1/2$ results from the uniform distribution of the phase noise without stabilization.
The derivation of $p_1$ is provided in Appendix~\ref{detailed_ana}.
In our protocol, we assume there is a deviation denoted as $d\phi$ between the phase difference of two successful SPI events and the deviation follows a normal distribution $p(d\phi)=\frac{1}{\sigma\sqrt{2\pi}}\exp((d\phi)^2/2\sigma^2)$ with a standard deviation $\sigma$.
Furthermore, the variance of $d\phi$ is assumed to be proportional to the fiber length $\sigma^2=DL$ with $D$ the diffusion coefficient~\cite{jiang2007fast,mina2008phasenoise}.
In this case, the average fidelity of entanglement from our protocol with respect to the maximal entanglement is $[1+\exp(-DL/2)]/2$.
With $p=0.07$~\cite{slodic2013atom}, $L_{\text{att}}=27.16$ km, $D=10^{-3}$ rad$^2$/km~\cite{jiang2007fast}, and $L=50$km, the fidelity of entanglement from SPI is less than 47.8\%.
In contrast, the fidelity of our protocol is more than 98\%.
Therefore, one can directly observe an evident advantage of our protocol in robustness against the phase instability.

\section{Discussion and conclusion}

Here, we briefly discuss a practical issue regarding the implementation of our quantum repeater scheme.
As stated in Sec.~\ref{qrep_protocol}, all memories generate and send photons to the intermediate station simultaneously with memory $j$ and $k$ sharing the same phase difference $\phi$.
As a result, our scheme can avoid the need for phase locking.
We point out that in practical scenarios some conditions must be met to leverage this advantage and reduce the phase stability requirement.
First, the frequencies of the generated photons at the same memory must be identical to achieve a successful SPI. 
The pump laser phase should be stabilized between two successful SPI events, ensuring that the phase difference of the pump laser remains consistent between these events.
Second, synchronizing photon emission from different memories is non-trivial.
A practical solution is to set a critical time interval within which phase fluctuations of the optical path can be neglected. 
If the interval between two successful SPI events exceeds this critical time, post-matching is not performed, and those events are discarded.
Finally, a spectrally multiplexed architecture can be employed to implement our scheme~\cite{sinclair2014spectral,grimau2017heralded}, where photons are coupled to the same spatial mode and share the same optical path fluctuation.
By meeting these conditions, the phase difference of the final distributed state at the elementary link in our protocol can be reduced by orders of magnitude compared with the DLCZ entanglement.
Therefore, even with realistic imperfections, our protocol can still significantly alleviate the phase stability requirement.

In summary, we introduce an asynchronous quantum repeater protocol that leverages the principle of post-matching.
We pair two successful SPI events and form a Bell entangled state with {\footnotesize CNOT} gate and $Z$-basis measurement.
Consequently, our protocol achieves the same entanglement efficiency as the quantum repeater protocol based on SPI.
Furthermore, our protocol is not constrained by the trade-off between the distribution rate and the fidelity with the maximal entanglement.
Additionally, our protocol alleviates the long-distance phase stability requirement, similar to protocols based on TPI.
Therefore, our protocol inherits the advantages of both SPI and TPI-based protocols.
We also present and analyze a specific implementation of our quantum repeater protocol based on the Kerr nonlinear resonator system, which shows the feasibility of our scheme with current technology.
One should note that, as we have shown in Table~\ref{comparison}, various physical systems are potential candidates for our quantum repeater protocol.
For instance, the trapped ion system can also be used to implement our scheme due to the ion's robust lifetime, long internal-state coherence, and strong ion-ion interactions~\cite{bruzewicz2019trapped}.
Single-qubit gates~\cite{brown2021single}, two-qubit gates~\cite{benhelm2008towards}, and qubit state preparation and readout~\cite{myerson2008high} have also been successfully performed using trapped ion system with high fidelity.

Given the benefits in communication efficiency and experiment mildness, we posit that a full functional quantum node, capable of both communication and information processing, can be constructed.
Therefore, our work represents a crucial step towards a fully-connected quantum network.
Our work also brings new insights about the connection between the quantum key distribution and quantum repeater with the idea of single-photon interference and post-matching. 
We anticipate this will benefit the development of the theory and technology in both fields.

$Note$ $added.$-After we post our work on arXiv, we notice a work from Yoshida \etal with a similar idea from arXiv~\cite{yoshida2024multiplexed}. This work introduces frequency multiplexing to the quantum repeater protocol and discusses implementing the protocol using rare-earth ion system.

We gratefully acknowledge the supports from the National Natural Science Foundation of China (No. 12274223), the Program for Innovative Talents and Entrepreneurs in Jiangsu (No. JSSCRC2021484), and the Fundamental Research Funds for the Central Universities and the Research Funds of Renmin University of China (No. 24XNKJ14).
C.-L.L. and H.-L.Y. contributed equally to this work.

\appendix

\section{Review of quantum repeater protocols}\label{review_rep}

In the main text, we compare the entanglement efficiency of our proposed protocol with that of protocols with entanglement generation based on SPI and TPI.
Therefore, here we briefly review these protocols.

\subsection{Quantum repeater protocol based on SPI}

\begin{figure}[tbp!]
		\includegraphics[width=8cm]{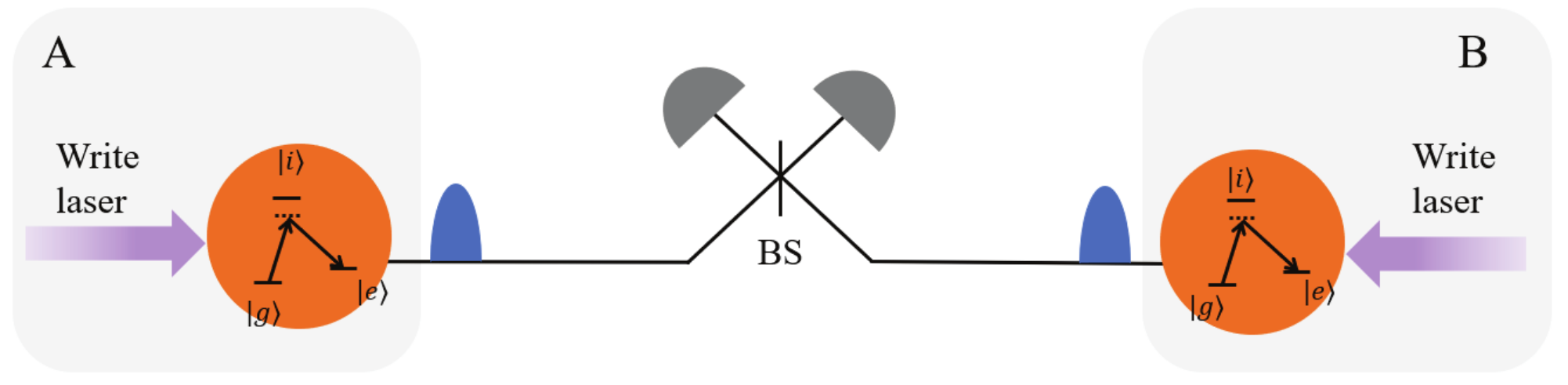}
		\caption{The schematic of the protocol based on SPI. The atomic ensembles or trapped ions located at both end nodes A and B are composed of three-level systems with a ground state $g$, a metastable state $e$, and an auxiliary state $i$, where at the beginning, all atoms are in the state $g$. A write laser leads to the spontaneous emission of the Stokes photons, which is then used to perform SPI. The generated Stokes photons are transmitted through quantum channel and combined on a beam splitter (BS) at an intermediate station.
		}\label{SPI}
\end{figure}

The entanglement generation based on SPI can be performed on various physical systems.
Here we take the DLCZ protocol which is based on the atomic ensembles as an example~\cite{duan2001long}.

The DLCZ protocol creates an atomic excitation-single photon entanglement using atomic ensembles and the photons can entangle two distant ensembles by SPI~\cite{duan2001long}.
To be specific, as shown in Fig.~\ref{SPI}, the atomic ensembles are composed of three-level systems with a ground state $g$, a metastable state $e$, and an auxiliary state $i$, where at the beginning, all atoms are in the state $g$.
A write laser leads to the spontaneous emission of the Stokes photons, which is then used to perform SPI.
In this process, the generation of two or more Stokes photons and the same number of atomic excitations are still possible, which is an important limiting factor in the quantum repeater protocol~\cite{duan2001long,sanguoard2011quantumrepeater}.

Now we present the principle of the creation of Bell entanglement at the elementary link.
The two ensembles located at two end nodes of one elementary link are simultaneously excited by a write laser and the state of the joint system is~\cite{sanguoard2011quantumrepeater}
\begin{equation}
    \left[1+\sqrt{\frac{p}{2}}(s^{\dagger}_aa^{\dagger}e^{i\phi_a}+s^{\dagger}_bb^{\dagger}e^{i\phi_b})+O(p)\right]\ket{0}.
\end{equation}
$a$ $(b)$ and $s_a$ ($s_b$) are annihilation operators of the Stokes photon and the atomic excitation associated with the corresponding ensembles.
$\phi_{a(b)}$ is the phase of the pump laser
.
$\frac{p}{2}$ is the probability of the emission of a single Stokes photon.
Here we only consider when there is only one Stokes photon and $O(p)$ represents the multiphoton terms as we have mentioned above.
The generated Stokes photons are transmitted through quantum channel and combined on a beam splitter at an intermediate station.
The above process can be described by the following state evolution (we denote the two outputs of the beam splitter as $d$ and $\tilde{d}$ and omit $O(p)$)
\begin{equation}\label{state_evo}
    \begin{split}
    &\left(1+e^{i\phi_a}\sqrt{\frac{p}{2}}(s_a^{\dagger}a^{\dagger}+e^{i\theta_{ab}}s_b^{\dagger}b^{\dagger})\right)\ket{0}\\
        &\overset{BS}{\rightarrow}\left(1+e^{i\phi_a}\sqrt{\frac{p}{4}}(s_a^{\dagger}(d^{\dagger}+\tilde{d}^{\dagger})+e^{i\theta_{ab}}s_b^{\dagger}(d^{\dagger}-\tilde{d}^{\dagger}))\right)\ket{0}\\
        &=\left(1+e^{i\phi_a}\sqrt{\frac{p}{4}}\left[(s_a^{\dagger}+e^{i\theta_{ab}}s_b^{\dagger})d^{\dagger}+(s_a^{\dagger}-e^{i\theta_{ab}}s_b^{\dagger})\tilde{d}^{\dagger}\right]\right)\ket{0}.
    \end{split}
\end{equation}

A successful SPI projects the state of the two atomic ensembles in
\begin{equation}\label{entangledSPI}
    \begin{split}
        \ket{\psi}_{ab}=&\frac{1}{\sqrt{2}}\left(s_a^{\dagger}\pm s_b^{\dagger}e^{i\theta_{ab}}\right)\ket{0}\\
        =&\frac{1}{\sqrt{2}}(\ket{10}_{ab}\pm e^{i\theta_{ab}}\ket{01}_{ab}).
    \end{split}
\end{equation}
$\ket{0}_{a(b)}$ denotes an ensemble with no atomic excitation, and $\ket{1}_{a(b)}$ denotes the storage of a single atomic excitation.
$\theta_{ab}$ is the phase difference.
From the state of the two ensembles, we can observe that a single atomic excitation is delocalized between two end nodes.
Considering the detection efficiency $\eta_d$ and the channel loss from one side to the intermediate station $\exp(-L/(2L_{att}))$, the probability of projecting the incoming state Eq.~(\ref{state_evo}) into the entangled state Eq.~(\ref{entangledSPI}) is given by $p\eta_d\exp(-L/(2L_{att}))$.
If we consider a trapped ion system where the entanglement between the atom and the photon is $(\sqrt{1-p}+\sqrt{p}s^{\dagger}a^{\dagger})\ket{0}$, a similar derivation can be made and the success probability of entanglement generation at the elementary link is $2p(1-p)\eta_d\exp(-L/(2L_{att}))$.

\subsection{Quantum repeater protocol based on TPI}

\begin{figure}[tbp!]
		\includegraphics[width=8cm]{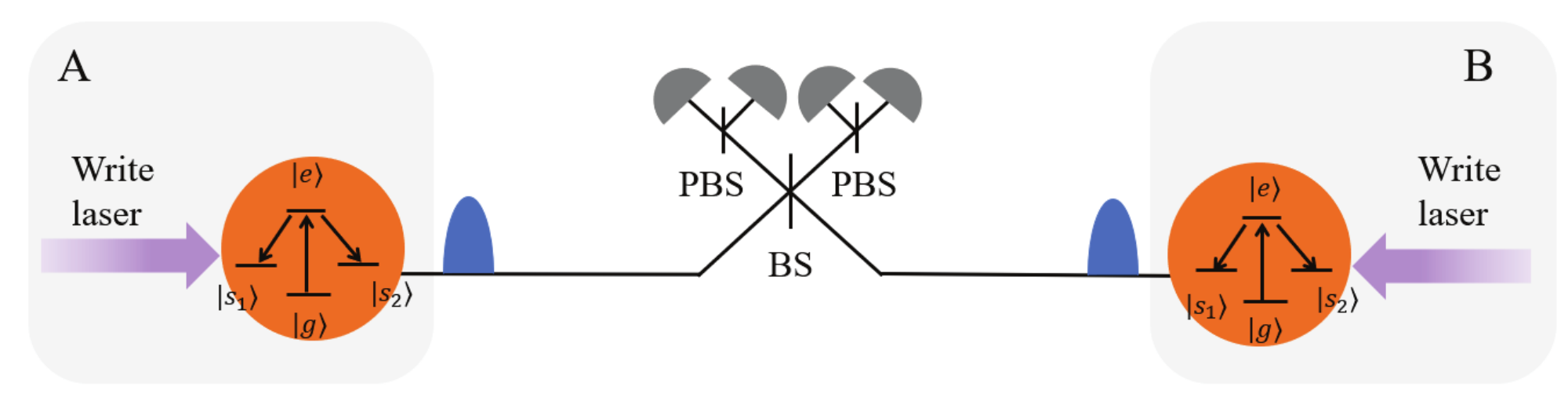}
		\caption{The schematic of the protocol based on TPI. At two end nodes, A and B each holds a trapped ion with a $\lambda$ system of energy levels, where the excited state $\ket{e}$ can decay into two degenerate metastable states $\ket{s_1}$ and $\ket{s_2}$ by emitting a photon into one of the two orthogonal polarization modes. The two end nodes both excite their ion into the excited state $\ket{e}_{A(B)}$ and the ion emits a single photon. The emitted photons propagate to an intermediate station where the TPI is performed by a Bell state analyzer composed of beam splitter (BS) and polarizing beam splitters (PBSs).
		}\label{TPI}
\end{figure}

Here we illustrate the scheme of entanglement generation based on TPI in the trapped ion system~\cite{simon2003robust}.

At two end nodes of an elementary link, each node holds a trapped ion with a $\lambda$ system of energy levels.
Specifically, the excited state $\ket{e}$ can decay into two degenerate metastable states $\ket{s_1}$ and $\ket{s_2}$ by emitting a photon into one of the two orthogonal polarization modes $a_1$ or $a_2$, where the probabilities of the transition from the exicted state to two metastable states are identical. 

Now we explain how the Bell entanglement is established at the elementary link.
The two end nodes both excite their ion into the excited state $\ket{e}_{A(B)}$.
When each ion emits a photon, the joint system of both ions and photons is 
\begin{equation}
    \frac{1}{2}\left(\ket{s_1}_Aa_1^{\dagger}+\ket{s_2}_Aa_2^{\dagger}\right)\left(\ket{s_1}_Bb_1^{\dagger}+\ket{s_2}_Bb_2^{\dagger}\right)\ket{0},
\end{equation}
where each ion is maximally entangled with the emitted photon.
The emitted photons propagate to an intermediate station where the TPI is performed.
The TPI is realized by Bell state analyzer composed of linear optical elements.
The Bell state analyzer can detect two Bell states $\ket{\psi^{\pm}}=(1/\sqrt{2})(a^{\dagger}_1b^{\dagger}_2\pm a^{\dagger}_2b^{\dagger}_1)\ket{0}$ out of four Bell states.
While the two remaining states $\ket{\phi^{\pm}}$ cannot be distinguished at the same time without non-linear optical elements.
The detection of the two photons in the state $\ket{\psi^{\pm}}$ projects the state of the two ions located at end nodes into the Bell state as well $(1/\sqrt{2})(\ket{s_1,s_2}_{AB}\pm\ket{s_2,s_1}_{AB})$, where the phases acquired by both photons are global.
Considering the detection imperfection $\eta_d$, the success probability of the TPI is $(1/2)\eta_d^2\exp(-L/L_{att})$.

\section{Pairing efficiency at the elementary link}\label{ent_eff_sec}

In this section, we derive the pairing efficiency $\Bar{n}/m$ with $\Bar{n}$ the average number of final Bell entanglement at the elementary link and $m$ the number of quantum memories.
We denote the success probability of SPI as $p_{\text{S}}$ and the probability of successful post-matching as $p_m$.
To be specific, $p_{\text{S}}$ can be calculated according to $2p(1-p)\eta_d\text{exp}(-L/(2L_{\text{att}}))$ as we have shown in the previous section.
In our simulation, we choose $p=0.07$~\cite{slodic2013atom}, $\eta_d=0.9$, and $L_{\text{att}}=27.16$ km.
The formula of $p_m$ is given in Sec.~\ref{evolution} of Appendix.

Now we present how to calculate $\Bar{n}$.
Before the {\footnotesize CNOT} operation and $Z$-basis measurement, $m$ SPIs are performed and the probability of obtaining $l$ pairs is
\begin{equation}\label{flm}
        f_{l|m}=B_{2l|m}(p_{\text{S}})+B_{2l+1|m}(p_{\text{S}})
\end{equation}
with $l=0,1,...,(m-1)/2$ when $m$ is odd and $l=0,1,...,m/2$ when $m$ is even.
$B_{k|m}(p)$ is the binomial distribution with
\begin{equation}
    B_{k|m}(p)=\begin{pmatrix}
        m\\k
    \end{pmatrix}p^k(1-p)^{m-k}.
\end{equation}
Note that when $m$ is even, we have $f_{\frac{m}{2}|m}=B_{m|m}(p_{\text{S}})=p_{\text{S}}^m$ which is inconsistent with Eq.~(\ref{flm}).
In order to obtain $l$ pairs, $2l$ or $2l+1$ successful SPI events are required and thus $f_{l|m}$ is the summation of $B_{2l|m}(p_{\text{S}})$ and $B_{2l+1|m}(p_{\text{S}})$.
After obtaining $l$ pairs, the {\footnotesize CNOT} operation and $Z$-basis measurement are performed and the probability of obtaining $n$ final Bell entanglement from $l$ pairs at the elementary link is $B_{n|l}(p_m)$.

Combining our results, we have the probability of obtaining $n$ Bell entanglement at the elementary link from $m$ SPIs
\begin{equation}
    P_{n|m}=\sum_{l=n}B_{n|l}(p_m)f_{l|m}.
\end{equation}
Correspondingly the average number of $n$ is 
\begin{equation}
    \Bar{n}=\sum_{n=0}nP_{n|m}=\sum_{n=0}n\sum_{l=n}B_{n|l}(p_m)f_{l|m}.
\end{equation}
When $m$ is odd, we have
\begin{equation}
    \begin{split}
        \Bar{n}=&\sum_{n=0}^{(m-1)/2}\sum_{l=n}^{(m-1)/2}nB_{n|l}(p_m)f_{l|m}\\
        =&\frac{m}{2}p_m[p_{\text{S}}-p_{\text{S}}(1-p_{\text{S}})^{m-1}]-\frac{1}{2}p_m\sum_{l=1}^{(m-1)/2}B_{2l+1|m}(p_{\text{S}}).
    \end{split}
\end{equation}
In the asymptotic regime ($m\rightarrow\infty$), we can conclude that
\begin{equation}\label{p0}
    \frac{\Bar{n}}{m}=\frac{1}{2}p_mp_{\text{S}}\sim\frac{1}{2}\exp(-L/(2L_{\text{att}})).
\end{equation}

\section{Calculating the average time required to establish entanglement over long distances}\label{avr_time}

In this section, we calculate the average time required to establish entangled pairs between two end nodes over long distance $L$.

We first calculate the average time $\tau(0,0)$ (given in units of $T_0$) for entanglement generation at the elementary link.
The probability of consuming $nT_0$ to generate Bell pair is given by the geometric distribution
\begin{equation}
    \text{P}(n,p_0)=(1-p_0)^{n-1}p_0,
\end{equation}
where $p_0$ is the success probability of generating Bell pair at the elementary link according to the proposed protocol and is given by Eq.~(\ref{p0}).
Then it is direct to calculate the average time $\tau(0,0)=\sum_n n\text{P}(n,p_0)=p_0^{-1}$. 

Then we derive a recurrence formula to calculate the average time $\tau(k_N,N)$.
We denote the success probability of entanglement distillation at $k_N$-the round of the $N$-th nesting level as $P_{\text{ED}}(k_N,N)$.
Following a similar argument as the above derivation, $\tau(k_N,N)$ is given by $P_{\text{ED}}^{-1}(k_N,N)$ in the units of the average time of obtaining two successful $(k_N-1)$-th round distillation events and classical communication.
In order to implement the $k_N$-th round distillation, two $(k_N-1)$-th round distillation events have to succeed.
According to Ref.~\cite{sanguoard2011quantumrepeater}, the average time of obtaining two successful $(k_N-1)$-th round distillation events is given by $(3/2)\tau(k_N-1,N)$ where the factor $3/2$ is a numerical approximation.
Apart from two pairs generated from the $(k_N-1)$-th round distillation, we need classical communication over $2^NL_0$ to acknowledge the result of the distillation.
Based on the above discussion, we have the following recurrence formula to calculate $\tau(k_N,N)$ $(k_N>0)$
\begin{equation}\label{kNN}
    \tau(k_N,N)=\frac{1}{P_{\text{ED}}(k_N,N)}\left[\frac{3}{2}\tau(k_N-1,N)+2^N\right].
\end{equation}

We will not stop here.
To solve the analytical expression of $\tau(k_N,N)$, we need another recurrence formula for $\tau(0,N)$ that is the average time needed before the first round distillation at the nesting level $N$.
As shown in the quantum repeater structure in the main text, before the first round distillation at the nesting level $N$, two $k_{N-1}$-th round distillation and the entanglement swapping have to succeed at the nesting level $N-1$.
Therefore, following the same derivation, we can give the recurrence formula
\begin{equation}\label{k0N}
    \tau(0,N)=\frac{1}{P_{\text{ES}}(N)}\left[\frac{3}{2}\tau(k_{N-1},N-1)+2^{N-1}\right]
\end{equation}
with $P_{\text{ES}}(N)$ the success probability of entanglement swapping at the nesting level $N$.
The factor $3/2$ exists for a similar reason as we discussed above.
$2^{N-1}$ given in units of $T_0$ is the classical communication time for acknowledging the success of swapping.

Now we can give the analytical formula of $\tau(k_N,N)$ by solving the recurrence relation.
According to Eq.~(\ref{kNN}), we have
\begin{equation}\label{kNN_sol}
    \begin{split}
        \tau(k_N,N)&=\tau(0,N)\left(\frac{3}{2}\right)^{k_N}\prod_{j=0}^{k_N-1}P^{-1}_{\text{ED}}(k_N-j,N)+\\
        &2^N\sum_{i=0}^{k_N-1}\left(\frac{3}{2}\right)^i\prod_{j=0}P^{-1}_{\text{ED}}(k_N-j,N)\\
        &=:\tau(0,N)\alpha(N)+\beta(N).
    \end{split}
\end{equation}
We insert Eq.~(\ref{k0N}) into Eq.~(\ref{kNN_sol}) and have
\begin{equation}
    \tau(k_N,N)=\frac{\alpha(N)}{P_{\text{ES}}(N)}\left(\frac{3}{2}\tau(k_{N-1},N-1)+2^{N-1}\right)+\beta(N).
\end{equation}
This is another recurrence formula and we can repeat the above process until
\begin{equation}
    \begin{split}
        \tau(k_N,N)&=\tau(k_0,0)\left(\frac{3}{2}\right)^N\prod_{j=0}^{N-1}\frac{\alpha(N-j)}{P_{\text{ES}}(N-j)}+\\
        &\sum_{i=1}^N\left(\frac{3}{2}\right)^{N-i}2^{i-1}\prod_{j=0}^{N-i}\frac{\alpha(N-j)}{P_{\text{ES}}(N-j)}+\\
        &\sum_{i=1}^N\left(\frac{3}{2}\right)^{N-i}\beta(i)\prod_{j=0}^{N-i-1}\frac{\alpha(N-j)}{P_{\text{ES}}(N-j)},
    \end{split}
\end{equation}
where $\tau(k_0,0)$ can be replaced by $\tau(0,0)\alpha(0)+\beta(0)$.

\section{Evolution of the entangled state under the noisy two-qubit gate}\label{evolution}

In this section, we present how the generated entanglement evolves during the quantum repeater protocol.

\subsection{Our protocol}
In our protocol, we transform two entangled states generated through SPI into a Bell state without phase difference by performing the {\footnotesize CNOT} operation and $Z$-basis measurement.
In this process, the imperfection of the {\footnotesize CNOT} gate will affect the success probability and the fidelity of the generated state.

We start with the state $\rho=\ket{\psi}_{AB}^{jk}\bra{\psi}$ of the joint system of memory $j$ and $k$ as given in the main text where SPIs are successfully performed.
After both end nodes applying the noisy {\footnotesize CNOT} gate on $\rho$, they measure the qubits $A$ and $B$ stored in memory $k$ on the $Z$ basis.
Ideally, the probability of the successful post-matching is $p_m=1/2$, $i.e.$ the success probability of both $A$ and $B$ being projected on \ket{\bar{1}}.
However, when taking the imperfect operation in consideration, this probability is transformed to $p_m=1/4+p_G/4$.
When $p_G=1$ ($i.$ $e.$ a perfect {\footnotesize CNOT} gate), the probability reduces to the ideal case.
If the post-matching is successful, the state of the generated noisy Bell pair at the elementary link can be given by the following density matrix
\begin{equation}\label{initial_state}
	\begin{pmatrix}
    \frac{1+3p_G}{4(1+p_G)} & 0 & 0 &\frac{p_G}{1+p_G} \\
    0 & \frac{1-p_G}{4(1+p_G)} & 0 & 0\\
    0 & 0 & \frac{1-p_G}{4(1+p_G)} & 0\\
    \frac{p_G}{1+p_G} & 0 & 0 & \frac{1+3p_G}{4(1+p_G)} 
\end{pmatrix}
\end{equation}
under the Hilbert space spanned by $\{\ket{gg},\ket{ge},\ket{eg},\ket{ee}\}$.
Note that we apply a $\sigma_x$ gate in calculating the matrix~(\ref{initial_state}) to obtain the Bell state $\ket{\phi^{+/-}}$ in the ideal case. 
This state is diagonal under the basis composed by the Bell states $\{\ket{\phi^+},\ket{\phi^-},\ket{\psi^+},\ket{\psi^-}\}$
\begin{equation}\label{initial_state_Bbas}
	\begin{pmatrix}
    \frac{1+7p_G}{4(1+p_G)} & 0 & 0 &0 \\
    0 & \frac{1-p_G}{4(1+p_G)} & 0 & 0\\
    0 & 0 & \frac{1-p_G}{4(1+p_G)} & 0\\
    0 & 0 & 0 & \frac{1-p_G}{4(1+p_G)} 
\end{pmatrix}.
\end{equation}
We write the generated state into the diagonal style to be fit for the following analysis considering the effect of entanglement distillation and entanglement swapping.

\subsection{Entanglement swapping}
Suppose the input state is $\rho_{ab}\otimes\rho_{cd}$.
The entanglement swapping start with applying a {\footnotesize CNOT} gate on $c$ conditioned on $b$.
Then one output system is measured in the computational basis $\{\ket{0},\ket{1}\}$ and the other one is measured in the basis $\{(\ket{0}+\ket{1})/\sqrt{2},(\ket{0}-\ket{1})/\sqrt{2}\}$.
Finally a single-qubit gate is performed depending on the measurement result.

Now we present the state after entanglement swapping.
If the input state is diagonal in the Bell basis, $i.$ $e.$ $\rho_{ab}=\rho_{cd}=A\ket{\phi^+}\bra{\phi^+}+B\ket{\phi^-}\bra{\phi^-}+C\ket{\psi^+}\bra{\psi^+}+D\ket{\psi^-}\bra{\psi^-}$, the resulting state of $a$ and $d$ is still diagonal with coefficients given by~\cite{abruzzo2013quantum}
\begin{equation}
    \begin{split}
        A'&=\frac{1-p_G}{4}+p_G(A^2+B^2+C^2+D^2),\\
        B'&=\frac{1-p_G}{4}+2p_G(AB+CD),\\
        C'&=\frac{1-p_G}{4}+2p_G(AC+BD),\\
        D'&=\frac{1-p_G}{4}+2p_G(AD+BC).\\
    \end{split}
\end{equation}
Note that here we consider an ideal detector and in this case the entanglement swapping is deterministic.

\subsection{Entanglement distillation}\label{ED}
We consider the $Oxford$ distillation protocol in our work~\cite{deutsch1996quantum}.
Similarly, when the input state is a Bell diagonal state, the output state of the distillation is still Bell diagonal with the following coefficients~\cite{abruzzo2013quantum}
\begin{widetext}
\begin{equation}\label{EDmap}
    \begin{split}
        A'&=\{1+p_G^2[(A-B-C+D)(3A+B\\
        &+C+3D)+4(A-D)^2]\}/(8P_{\text{ED}}(k_N,N)),\\
        B'&=\{1-p_G^2[A^2+2A(B+C-7D)\\
        &+(B+C+D)^2]\}/(8P_{\text{ED}}(k_N,N)),\\
        C'&=\{1+p_G^2[4(B-C)^2-(A-B-C+D)\\
        &(A+3B+3C+D)]\}/(8P_{\text{ED}}(k_N,N)),\\
        D'&=\{1-p_G^2[A^2+2A(B+C+D)+B^2\\
        &+2B(D-7C)+(C+D)^2]\}/(8P_{\text{ED}}(k_N,N)).\\
    \end{split}
\end{equation}
\end{widetext}
$P_{\text{ED}}(k_N,N)$ is the success probability of the $k_N$-th round distillation at the $N$-th nesting level, where $P_{\text{ED}}(k_N,N)=(1/2)[1+p_G^2(2A_{k_N-1}+2D_{k_N-1}-1)^2]$.

\section{Detailed analysis of state evolution in our protocol}\label{detailed_ana}

In this section, we provide a detailed analysis of the state evolution in our protocol taking the higher excitation term into consideration.

In the following, we take into account the channel loss and assume a photon number resolving detector with unit efficiency.
Our protocol starts with the qubit-photon entangled states from A and B. 
Then qubits are retained while photons are transmitted to the central station to perform single-photon interference.
Conditioned on detecting one photon at one output mode and no photon at the other, the density operator between qubits held by node A and B can be given by 
\begin{equation}\label{noisy_ent}
    \rho_{10}=p_{1}\ket{\psi^+(\theta)}_{AB}\bra{\psi^+(\theta)}+(1-p_1)\ket{11}_{AB}\bra{11},
\end{equation}
where $\ket{\psi^+(\theta)}=(\ket{01}+e^{i\theta}\ket{10})/\sqrt{2}$, $p_1 = p(1-p)\eta_t/P_{10}$, and $P_{10}=p(1-p)\eta_t+p^2(1-\eta_t)\eta_t$.
$\theta$ is the phase difference and $p$ is the probability of generating one photon in the qubit-photon entanglement as defined in the main text.
$\eta_t=\text{exp}[-L/(2L_{\text{att}})]$ is the channel loss from one end node to the central station.
The subscript 10 of the density operator means the detection of one photon at output mode $d$ and no photon at mode $\tilde{d}$.
The additional term $\ket{11}\bra{11}$ appears since two photons arrive at the central station and only one of them is detected.

Now we consider the state evolution during post-matching, factoring in the imperfect {\footnotesize CNOT} operation.
We assume two pairs of noisy entangled state are used for post-matching, both established under the condition that one photon is detected at output mode $d$ and no photon at mode $\tilde{d}$.
The state of the joint system of two pairs of noisy entangled state is given by 
\begin{widetext}
\begin{equation}
    \begin{split}
        \rho^j_{10}\otimes\rho^k_{10}=&p_1^2\ket{\psi^+(\theta)}_j\bra{\psi^+(\theta)}\otimes\ket{\psi^+(\theta)}_k\bra{\psi^+(\theta)}+p_1(1-p_1)\ket{\psi^+(\theta)}_j\bra{\psi^+(\theta)}\otimes\ket{11}_k\bra{11}+\\
        &(1-p_1)p_1\ket{11}_j\bra{11}\otimes\ket{\psi^+(\theta)}_k\bra{\psi^+(\theta)}+(1-p_1)^2\ket{11}_j\bra{11}\otimes\ket{11}_k\bra{11}.
    \end{split}
\end{equation}
\end{widetext}

During the post-matching, a {\footnotesize CNOT} operation is first applied at nodes A and B with memory $j$ as the condition qubit and memory $k$ as the target qubit.
Then qubits in memory $k$ are then measured in the $Z$ basis and both nodes keep the qubits in memory $j$ when the result of projection is $\ket{11}$.
After the projection, the state of the system is given by a maximally entangled state without phase difference
\begin{equation}
    \rho_{10,\text{ele}}=\ket{\psi^+(0)}\bra{\psi^+(0)}
\end{equation}
with success probability $p^2_1/2$, where 1/2 accounts for the success probability of $Z$ projection on $\ket{11}$.
The subscript ele stands for the entanglement in elementary link.

We offer a brief remark here.
The post-matching in our protocol consisting of {\footnotesize CNOT} operation and $Z$ measurement can actually be regarded as an inherent entanglement purification.
Compared to $\rho_{10}$, the term $\ket{11}\bra{11}$ term and the phase difference $\theta$ in $\rho_{10,\text{ele}}$ vanishes, resulting in a maximally entangled state between nodes A and B, as indicated by the expressions above.
With an ideal {\footnotesize CNOT} operation ($p_G=1$), the term $\ket{11}\bra{11}$ vanishes, resulting in a maximally entangled state between nodes A and B.
Since the entanglement generated from the single-photon interference always follows a similar expression as $\rho_{10}$ with an additional term $\ket{11}\bra{11}$, the post-matching operation in our protocol can always distill a maximally entangled state regardless of the initial visibility.
Furthermore, the fidelity of the final distributed state with respect to $\ket{\psi^+}$ is always 1 after entanglement distillation operation in post-matching of our protocol.

%\bibliography{ref3}

\begin{thebibliography}{66}%
\makeatletter
\providecommand \@ifxundefined [1]{%
 \@ifx{#1\undefined}
}%
\providecommand \@ifnum [1]{%
 \ifnum #1\expandafter \@firstoftwo
 \else \expandafter \@secondoftwo
 \fi
}%
\providecommand \@ifx [1]{%
 \ifx #1\expandafter \@firstoftwo
 \else \expandafter \@secondoftwo
 \fi
}%
\providecommand \natexlab [1]{#1}%
\providecommand \enquote  [1]{``#1''}%
\providecommand \bibnamefont  [1]{#1}%
\providecommand \bibfnamefont [1]{#1}%
\providecommand \citenamefont [1]{#1}%
\providecommand \href@noop [0]{\@secondoftwo}%
\providecommand \href [0]{\begingroup \@sanitize@url \@href}%
\providecommand \@href[1]{\@@startlink{#1}\@@href}%
\providecommand \@@href[1]{\endgroup#1\@@endlink}%
\providecommand \@sanitize@url [0]{\catcode `\\12\catcode `\$12\catcode `\&12\catcode `\#12\catcode `\^12\catcode `\_12\catcode `\%12\relax}%
\providecommand \@@startlink[1]{}%
\providecommand \@@endlink[0]{}%
\providecommand \url  [0]{\begingroup\@sanitize@url \@url }%
\providecommand \@url [1]{\endgroup\@href {#1}{\urlprefix }}%
\providecommand \urlprefix  [0]{URL }%
\providecommand \Eprint [0]{\href }%
\providecommand \doibase [0]{https://doi.org/}%
\providecommand \selectlanguage [0]{\@gobble}%
\providecommand \bibinfo  [0]{\@secondoftwo}%
\providecommand \bibfield  [0]{\@secondoftwo}%
\providecommand \translation [1]{[#1]}%
\providecommand \BibitemOpen [0]{}%
\providecommand \bibitemStop [0]{}%
\providecommand \bibitemNoStop [0]{.\EOS\space}%
\providecommand \EOS [0]{\spacefactor3000\relax}%
\providecommand \BibitemShut  [1]{\csname bibitem#1\endcsname}%
\let\auto@bib@innerbib\@empty
%</preamble>
\bibitem [{\citenamefont {Bennett}\ and\ \citenamefont {Wiesner}(1992)}]{bennett1992communications}%
  \BibitemOpen
  \bibfield  {author} {\bibinfo {author} {\bibfnamefont {C.~H.}\ \bibnamefont {Bennett}}\ and\ \bibinfo {author} {\bibfnamefont {S.~J.}\ \bibnamefont {Wiesner}},\ }\bibfield  {title} {\bibinfo {title} {Communication via one- and two-particle operators on {Einstein-Podolsky-Rosen} states},\ }\href@noop {} {\bibfield  {journal} {\bibinfo  {journal} {Phys. Rev. Lett.}\ }\textbf {\bibinfo {volume} {69}},\ \bibinfo {pages} {2881} (\bibinfo {year} {1992})}\BibitemShut {NoStop}%
\bibitem [{\citenamefont {Xie}\ \emph {et~al.}(2022)\citenamefont {Xie}, \citenamefont {Lu}, \citenamefont {Weng}, \citenamefont {Cao}, \citenamefont {Jia}, \citenamefont {Bao}, \citenamefont {Wang}, \citenamefont {Fu}, \citenamefont {Yin},\ and\ \citenamefont {Chen}}]{xie2022breaking}%
  \BibitemOpen
  \bibfield  {author} {\bibinfo {author} {\bibfnamefont {Y.-M.}\ \bibnamefont {Xie}}, \bibinfo {author} {\bibfnamefont {Y.-S.}\ \bibnamefont {Lu}}, \bibinfo {author} {\bibfnamefont {C.-X.}\ \bibnamefont {Weng}}, \bibinfo {author} {\bibfnamefont {X.-Y.}\ \bibnamefont {Cao}}, \bibinfo {author} {\bibfnamefont {Z.-Y.}\ \bibnamefont {Jia}}, \bibinfo {author} {\bibfnamefont {Y.}~\bibnamefont {Bao}}, \bibinfo {author} {\bibfnamefont {Y.}~\bibnamefont {Wang}}, \bibinfo {author} {\bibfnamefont {Y.}~\bibnamefont {Fu}}, \bibinfo {author} {\bibfnamefont {H.-L.}\ \bibnamefont {Yin}},\ and\ \bibinfo {author} {\bibfnamefont {Z.-B.}\ \bibnamefont {Chen}},\ }\bibfield  {title} {\bibinfo {title} {Breaking the rate-loss bound of quantum key distribution with asynchronous two-photon interference},\ }\href@noop {} {\bibfield  {journal} {\bibinfo  {journal} {PRX Quantum}\ }\textbf {\bibinfo {volume} {3}},\ \bibinfo {pages} {020315} (\bibinfo {year} {2022})}\BibitemShut {NoStop}%
\bibitem [{\citenamefont {Illiano}\ \emph {et~al.}(2022)\citenamefont {Illiano}, \citenamefont {Caleffi}, \citenamefont {Manzalini},\ and\ \citenamefont {Cacciapuoti}}]{jessica2022quantum}%
  \BibitemOpen
  \bibfield  {author} {\bibinfo {author} {\bibfnamefont {J.}~\bibnamefont {Illiano}}, \bibinfo {author} {\bibfnamefont {M.}~\bibnamefont {Caleffi}}, \bibinfo {author} {\bibfnamefont {A.}~\bibnamefont {Manzalini}},\ and\ \bibinfo {author} {\bibfnamefont {A.~S.}\ \bibnamefont {Cacciapuoti}},\ }\bibfield  {title} {\bibinfo {title} {Quantum internet protocol stack: A comprehensive survey},\ }\href@noop {} {\bibfield  {journal} {\bibinfo  {journal} {Computer Networks}\ }\textbf {\bibinfo {volume} {213}},\ \bibinfo {pages} {109092} (\bibinfo {year} {2022})}\BibitemShut {NoStop}%
\bibitem [{\citenamefont {Arute}\ \emph {et~al.}(2019)\citenamefont {Arute}, \citenamefont {Arya}, \citenamefont {Babbush}, \citenamefont {Bacon}, \citenamefont {Bardin}, \citenamefont {Barends}, \citenamefont {Biswas}, \citenamefont {Boixo}, \citenamefont {Brandao}, \citenamefont {Buell} \emph {et~al.}}]{arute2019quantum}%
  \BibitemOpen
  \bibfield  {author} {\bibinfo {author} {\bibfnamefont {F.}~\bibnamefont {Arute}}, \bibinfo {author} {\bibfnamefont {K.}~\bibnamefont {Arya}}, \bibinfo {author} {\bibfnamefont {R.}~\bibnamefont {Babbush}}, \bibinfo {author} {\bibfnamefont {D.}~\bibnamefont {Bacon}}, \bibinfo {author} {\bibfnamefont {J.~C.}\ \bibnamefont {Bardin}}, \bibinfo {author} {\bibfnamefont {R.}~\bibnamefont {Barends}}, \bibinfo {author} {\bibfnamefont {R.}~\bibnamefont {Biswas}}, \bibinfo {author} {\bibfnamefont {S.}~\bibnamefont {Boixo}}, \bibinfo {author} {\bibfnamefont {F.~G.}\ \bibnamefont {Brandao}}, \bibinfo {author} {\bibfnamefont {D.~A.}\ \bibnamefont {Buell}}, \emph {et~al.},\ }\bibfield  {title} {\bibinfo {title} {Quantum supremacy using a programmable superconducting processor},\ }\href@noop {} {\bibfield  {journal} {\bibinfo  {journal} {Nature}\ }\textbf {\bibinfo {volume} {574}},\ \bibinfo {pages} {505} (\bibinfo {year} {2019})}\BibitemShut {NoStop}%
\bibitem [{\citenamefont {Zhong}\ \emph {et~al.}(2020)\citenamefont {Zhong}, \citenamefont {Wang}, \citenamefont {Deng}, \citenamefont {Chen}, \citenamefont {Peng}, \citenamefont {Luo}, \citenamefont {Qin}, \citenamefont {Wu}, \citenamefont {Ding}, \citenamefont {Hu} \emph {et~al.}}]{zhong2020quantum}%
  \BibitemOpen
  \bibfield  {author} {\bibinfo {author} {\bibfnamefont {H.-S.}\ \bibnamefont {Zhong}}, \bibinfo {author} {\bibfnamefont {H.}~\bibnamefont {Wang}}, \bibinfo {author} {\bibfnamefont {Y.-H.}\ \bibnamefont {Deng}}, \bibinfo {author} {\bibfnamefont {M.-C.}\ \bibnamefont {Chen}}, \bibinfo {author} {\bibfnamefont {L.-C.}\ \bibnamefont {Peng}}, \bibinfo {author} {\bibfnamefont {Y.-H.}\ \bibnamefont {Luo}}, \bibinfo {author} {\bibfnamefont {J.}~\bibnamefont {Qin}}, \bibinfo {author} {\bibfnamefont {D.}~\bibnamefont {Wu}}, \bibinfo {author} {\bibfnamefont {X.}~\bibnamefont {Ding}}, \bibinfo {author} {\bibfnamefont {Y.}~\bibnamefont {Hu}}, \emph {et~al.},\ }\bibfield  {title} {\bibinfo {title} {Quantum computational advantage using photons},\ }\href@noop {} {\bibfield  {journal} {\bibinfo  {journal} {Science}\ }\textbf {\bibinfo {volume} {370}},\ \bibinfo {pages} {1460} (\bibinfo {year} {2020})}\BibitemShut {NoStop}%
\bibitem [{\citenamefont {Zhou}\ \emph {et~al.}(2022)\citenamefont {Zhou}, \citenamefont {Cao}, \citenamefont {Lu}, \citenamefont {Wang}, \citenamefont {Bao}, \citenamefont {Jia}, \citenamefont {Fu}, \citenamefont {Yin},\ and\ \citenamefont {Chen}}]{zhou2022experimental}%
  \BibitemOpen
  \bibfield  {author} {\bibinfo {author} {\bibfnamefont {M.-G.}\ \bibnamefont {Zhou}}, \bibinfo {author} {\bibfnamefont {X.-Y.}\ \bibnamefont {Cao}}, \bibinfo {author} {\bibfnamefont {Y.-S.}\ \bibnamefont {Lu}}, \bibinfo {author} {\bibfnamefont {Y.}~\bibnamefont {Wang}}, \bibinfo {author} {\bibfnamefont {Y.}~\bibnamefont {Bao}}, \bibinfo {author} {\bibfnamefont {Z.-Y.}\ \bibnamefont {Jia}}, \bibinfo {author} {\bibfnamefont {Y.}~\bibnamefont {Fu}}, \bibinfo {author} {\bibfnamefont {H.-L.}\ \bibnamefont {Yin}},\ and\ \bibinfo {author} {\bibfnamefont {Z.-B.}\ \bibnamefont {Chen}},\ }\bibfield  {title} {\bibinfo {title} {Experimental quantum advantage with quantum coupon collector},\ }\href@noop {} {\bibfield  {journal} {\bibinfo  {journal} {Research}\ }\textbf {\bibinfo {volume} {2022}},\ \bibinfo {pages} {9798679} (\bibinfo {year} {2022})}\BibitemShut {NoStop}%
\bibitem [{\citenamefont {Cacciapuoti}\ \emph {et~al.}(2020)\citenamefont {Cacciapuoti}, \citenamefont {Caleffi}, \citenamefont {Tafuri}, \citenamefont {Cataliotti}, \citenamefont {Gherardini},\ and\ \citenamefont {Bianchi}}]{caccoapuoti2020quantum}%
  \BibitemOpen
  \bibfield  {author} {\bibinfo {author} {\bibfnamefont {A.~S.}\ \bibnamefont {Cacciapuoti}}, \bibinfo {author} {\bibfnamefont {M.}~\bibnamefont {Caleffi}}, \bibinfo {author} {\bibfnamefont {F.}~\bibnamefont {Tafuri}}, \bibinfo {author} {\bibfnamefont {F.~S.}\ \bibnamefont {Cataliotti}}, \bibinfo {author} {\bibfnamefont {S.}~\bibnamefont {Gherardini}},\ and\ \bibinfo {author} {\bibfnamefont {G.}~\bibnamefont {Bianchi}},\ }\bibfield  {title} {\bibinfo {title} {Quantum internet: Networking challenges in distributed quantum computing},\ }\href@noop {} {\bibfield  {journal} {\bibinfo  {journal} {IEEE Network}\ }\textbf {\bibinfo {volume} {34}},\ \bibinfo {pages} {137} (\bibinfo {year} {2020})}\BibitemShut {NoStop}%
\bibitem [{\citenamefont {Duan}\ \emph {et~al.}(2001)\citenamefont {Duan}, \citenamefont {Lukin}, \citenamefont {Cirac},\ and\ \citenamefont {Zoller}}]{duan2001long}%
  \BibitemOpen
  \bibfield  {author} {\bibinfo {author} {\bibfnamefont {L.-M.}\ \bibnamefont {Duan}}, \bibinfo {author} {\bibfnamefont {M.~D.}\ \bibnamefont {Lukin}}, \bibinfo {author} {\bibfnamefont {J.~I.}\ \bibnamefont {Cirac}},\ and\ \bibinfo {author} {\bibfnamefont {P.}~\bibnamefont {Zoller}},\ }\bibfield  {title} {\bibinfo {title} {Long-distance quantum communication with atomic ensembles and linear optics},\ }\href@noop {} {\bibfield  {journal} {\bibinfo  {journal} {Nature}\ }\textbf {\bibinfo {volume} {414}},\ \bibinfo {pages} {413} (\bibinfo {year} {2001})}\BibitemShut {NoStop}%
\bibitem [{\citenamefont {Zhao}\ \emph {et~al.}(2007)\citenamefont {Zhao}, \citenamefont {Chen}, \citenamefont {Chen}, \citenamefont {Schmiedmayer},\ and\ \citenamefont {Pan}}]{zhao2007robust}%
  \BibitemOpen
  \bibfield  {author} {\bibinfo {author} {\bibfnamefont {B.}~\bibnamefont {Zhao}}, \bibinfo {author} {\bibfnamefont {Z.-B.}\ \bibnamefont {Chen}}, \bibinfo {author} {\bibfnamefont {Y.-A.}\ \bibnamefont {Chen}}, \bibinfo {author} {\bibfnamefont {J.}~\bibnamefont {Schmiedmayer}},\ and\ \bibinfo {author} {\bibfnamefont {J.-W.}\ \bibnamefont {Pan}},\ }\bibfield  {title} {\bibinfo {title} {Robust creation of entanglement between remote memory qubits},\ }\href@noop {} {\bibfield  {journal} {\bibinfo  {journal} {Phys. Rev. Lett.}\ }\textbf {\bibinfo {volume} {98}},\ \bibinfo {pages} {240502} (\bibinfo {year} {2007})}\BibitemShut {NoStop}%
\bibitem [{\citenamefont {Sangouard}\ \emph {et~al.}(2011)\citenamefont {Sangouard}, \citenamefont {Simon}, \citenamefont {de~Riedmatten},\ and\ \citenamefont {Gisin}}]{sanguoard2011quantumrepeater}%
  \BibitemOpen
  \bibfield  {author} {\bibinfo {author} {\bibfnamefont {N.}~\bibnamefont {Sangouard}}, \bibinfo {author} {\bibfnamefont {C.}~\bibnamefont {Simon}}, \bibinfo {author} {\bibfnamefont {H.}~\bibnamefont {de~Riedmatten}},\ and\ \bibinfo {author} {\bibfnamefont {N.}~\bibnamefont {Gisin}},\ }\bibfield  {title} {\bibinfo {title} {Quantum repeaters based on atomic ensembles and linear optics},\ }\href@noop {} {\bibfield  {journal} {\bibinfo  {journal} {Rev. Mod. Phys.}\ }\textbf {\bibinfo {volume} {83}},\ \bibinfo {pages} {33} (\bibinfo {year} {2011})}\BibitemShut {NoStop}%
\bibitem [{\citenamefont {Azuma}\ \emph {et~al.}(2023)\citenamefont {Azuma}, \citenamefont {Economou}, \citenamefont {Elkouss}, \citenamefont {Hilaire}, \citenamefont {Jiang}, \citenamefont {Lo},\ and\ \citenamefont {Tzitrin}}]{Azuma2023Quantum}%
  \BibitemOpen
  \bibfield  {author} {\bibinfo {author} {\bibfnamefont {K.}~\bibnamefont {Azuma}}, \bibinfo {author} {\bibfnamefont {S.~E.}\ \bibnamefont {Economou}}, \bibinfo {author} {\bibfnamefont {D.}~\bibnamefont {Elkouss}}, \bibinfo {author} {\bibfnamefont {P.}~\bibnamefont {Hilaire}}, \bibinfo {author} {\bibfnamefont {L.}~\bibnamefont {Jiang}}, \bibinfo {author} {\bibfnamefont {H.-K.}\ \bibnamefont {Lo}},\ and\ \bibinfo {author} {\bibfnamefont {I.}~\bibnamefont {Tzitrin}},\ }\bibfield  {title} {\bibinfo {title} {Quantum repeaters: From quantum networks to the quantum internet},\ }\href@noop {} {\bibfield  {journal} {\bibinfo  {journal} {Rev. Mod. Phys.}\ }\textbf {\bibinfo {volume} {95}},\ \bibinfo {pages} {045006} (\bibinfo {year} {2023})}\BibitemShut {NoStop}%
\bibitem [{\citenamefont {Munro}\ \emph {et~al.}(2010)\citenamefont {Munro}, \citenamefont {Harrison}, \citenamefont {Stephens}, \citenamefont {Devitt},\ and\ \citenamefont {Nemoto}}]{munro2010quantum}%
  \BibitemOpen
  \bibfield  {author} {\bibinfo {author} {\bibfnamefont {W.}~\bibnamefont {Munro}}, \bibinfo {author} {\bibfnamefont {K.}~\bibnamefont {Harrison}}, \bibinfo {author} {\bibfnamefont {A.}~\bibnamefont {Stephens}}, \bibinfo {author} {\bibfnamefont {S.}~\bibnamefont {Devitt}},\ and\ \bibinfo {author} {\bibfnamefont {K.}~\bibnamefont {Nemoto}},\ }\bibfield  {title} {\bibinfo {title} {From quantum multiplexing to high-performance quantum networking},\ }\href@noop {} {\bibfield  {journal} {\bibinfo  {journal} {Nat. Photonics}\ }\textbf {\bibinfo {volume} {4}},\ \bibinfo {pages} {792} (\bibinfo {year} {2010})}\BibitemShut {NoStop}%
\bibitem [{\citenamefont {Munro}\ \emph {et~al.}(2012)\citenamefont {Munro}, \citenamefont {Stephens}, \citenamefont {Devitt}, \citenamefont {Harrison},\ and\ \citenamefont {Nemoto}}]{munro2012quantum}%
  \BibitemOpen
  \bibfield  {author} {\bibinfo {author} {\bibfnamefont {W.~J.}\ \bibnamefont {Munro}}, \bibinfo {author} {\bibfnamefont {A.~M.}\ \bibnamefont {Stephens}}, \bibinfo {author} {\bibfnamefont {S.~J.}\ \bibnamefont {Devitt}}, \bibinfo {author} {\bibfnamefont {K.~A.}\ \bibnamefont {Harrison}},\ and\ \bibinfo {author} {\bibfnamefont {K.}~\bibnamefont {Nemoto}},\ }\bibfield  {title} {\bibinfo {title} {Quantum communication without the necessity of quantum memories},\ }\href@noop {} {\bibfield  {journal} {\bibinfo  {journal} {Nat. Photonics}\ }\textbf {\bibinfo {volume} {6}},\ \bibinfo {pages} {777} (\bibinfo {year} {2012})}\BibitemShut {NoStop}%
\bibitem [{\citenamefont {Fowler}\ \emph {et~al.}(2010)\citenamefont {Fowler}, \citenamefont {Wang}, \citenamefont {Hill}, \citenamefont {Ladd}, \citenamefont {Van~Meter},\ and\ \citenamefont {Hollenberg}}]{fowler2010surface}%
  \BibitemOpen
  \bibfield  {author} {\bibinfo {author} {\bibfnamefont {A.~G.}\ \bibnamefont {Fowler}}, \bibinfo {author} {\bibfnamefont {D.~S.}\ \bibnamefont {Wang}}, \bibinfo {author} {\bibfnamefont {C.~D.}\ \bibnamefont {Hill}}, \bibinfo {author} {\bibfnamefont {T.~D.}\ \bibnamefont {Ladd}}, \bibinfo {author} {\bibfnamefont {R.}~\bibnamefont {Van~Meter}},\ and\ \bibinfo {author} {\bibfnamefont {L.~C.~L.}\ \bibnamefont {Hollenberg}},\ }\bibfield  {title} {\bibinfo {title} {Surface code quantum communication},\ }\href@noop {} {\bibfield  {journal} {\bibinfo  {journal} {Phys. Rev. Lett.}\ }\textbf {\bibinfo {volume} {104}},\ \bibinfo {pages} {180503} (\bibinfo {year} {2010})}\BibitemShut {NoStop}%
\bibitem [{\citenamefont {Muralidharan}\ \emph {et~al.}(2014)\citenamefont {Muralidharan}, \citenamefont {Kim}, \citenamefont {L\"utkenhaus}, \citenamefont {Lukin},\ and\ \citenamefont {Jiang}}]{muralidharan2014ultrafast}%
  \BibitemOpen
  \bibfield  {author} {\bibinfo {author} {\bibfnamefont {S.}~\bibnamefont {Muralidharan}}, \bibinfo {author} {\bibfnamefont {J.}~\bibnamefont {Kim}}, \bibinfo {author} {\bibfnamefont {N.}~\bibnamefont {L\"utkenhaus}}, \bibinfo {author} {\bibfnamefont {M.~D.}\ \bibnamefont {Lukin}},\ and\ \bibinfo {author} {\bibfnamefont {L.}~\bibnamefont {Jiang}},\ }\bibfield  {title} {\bibinfo {title} {Ultrafast and fault-tolerant quantum communication across long distances},\ }\href@noop {} {\bibfield  {journal} {\bibinfo  {journal} {Phys. Rev. Lett.}\ }\textbf {\bibinfo {volume} {112}},\ \bibinfo {pages} {250501} (\bibinfo {year} {2014})}\BibitemShut {NoStop}%
\bibitem [{\citenamefont {Azuma}\ \emph {et~al.}(2015)\citenamefont {Azuma}, \citenamefont {Tamaki},\ and\ \citenamefont {Lo}}]{azuma2015all}%
  \BibitemOpen
  \bibfield  {author} {\bibinfo {author} {\bibfnamefont {K.}~\bibnamefont {Azuma}}, \bibinfo {author} {\bibfnamefont {K.}~\bibnamefont {Tamaki}},\ and\ \bibinfo {author} {\bibfnamefont {H.-K.}\ \bibnamefont {Lo}},\ }\bibfield  {title} {\bibinfo {title} {All-photonic quantum repeaters},\ }\href@noop {} {\bibfield  {journal} {\bibinfo  {journal} {Nat. Commun.}\ }\textbf {\bibinfo {volume} {6}},\ \bibinfo {pages} {6787} (\bibinfo {year} {2015})}\BibitemShut {NoStop}%
\bibitem [{\citenamefont {Li}\ \emph {et~al.}(2023)\citenamefont {Li}, \citenamefont {Fu}, \citenamefont {Liu}, \citenamefont {Xie}, \citenamefont {Li}, \citenamefont {Zhou}, \citenamefont {Yin},\ and\ \citenamefont {Chen}}]{li2023alphotonic}%
  \BibitemOpen
  \bibfield  {author} {\bibinfo {author} {\bibfnamefont {C.-L.}\ \bibnamefont {Li}}, \bibinfo {author} {\bibfnamefont {Y.}~\bibnamefont {Fu}}, \bibinfo {author} {\bibfnamefont {W.-B.}\ \bibnamefont {Liu}}, \bibinfo {author} {\bibfnamefont {Y.-M.}\ \bibnamefont {Xie}}, \bibinfo {author} {\bibfnamefont {B.-H.}\ \bibnamefont {Li}}, \bibinfo {author} {\bibfnamefont {M.-G.}\ \bibnamefont {Zhou}}, \bibinfo {author} {\bibfnamefont {H.-L.}\ \bibnamefont {Yin}},\ and\ \bibinfo {author} {\bibfnamefont {Z.-B.}\ \bibnamefont {Chen}},\ }\bibfield  {title} {\bibinfo {title} {All-photonic quantum repeater for multipartite entanglement generation},\ }\href@noop {} {\bibfield  {journal} {\bibinfo  {journal} {Opt. Lett.}\ }\textbf {\bibinfo {volume} {48}},\ \bibinfo {pages} {1244} (\bibinfo {year} {2023})}\BibitemShut {NoStop}%
\bibitem [{\citenamefont {Yu}\ \emph {et~al.}(2020)\citenamefont {Yu}, \citenamefont {Ma}, \citenamefont {Luo}, \citenamefont {Jing}, \citenamefont {Sun}, \citenamefont {Fang}, \citenamefont {Yang}, \citenamefont {Liu}, \citenamefont {Zheng}, \citenamefont {Xie} \emph {et~al.}}]{yu2020entanglement}%
  \BibitemOpen
  \bibfield  {author} {\bibinfo {author} {\bibfnamefont {Y.}~\bibnamefont {Yu}}, \bibinfo {author} {\bibfnamefont {F.}~\bibnamefont {Ma}}, \bibinfo {author} {\bibfnamefont {X.-Y.}\ \bibnamefont {Luo}}, \bibinfo {author} {\bibfnamefont {B.}~\bibnamefont {Jing}}, \bibinfo {author} {\bibfnamefont {P.-F.}\ \bibnamefont {Sun}}, \bibinfo {author} {\bibfnamefont {R.-Z.}\ \bibnamefont {Fang}}, \bibinfo {author} {\bibfnamefont {C.-W.}\ \bibnamefont {Yang}}, \bibinfo {author} {\bibfnamefont {H.}~\bibnamefont {Liu}}, \bibinfo {author} {\bibfnamefont {M.-Y.}\ \bibnamefont {Zheng}}, \bibinfo {author} {\bibfnamefont {X.-P.}\ \bibnamefont {Xie}}, \emph {et~al.},\ }\bibfield  {title} {\bibinfo {title} {Entanglement of two quantum memories via fibres over dozens of kilometres},\ }\href@noop {} {\bibfield  {journal} {\bibinfo  {journal} {Nature}\ }\textbf {\bibinfo {volume} {578}},\ \bibinfo {pages} {240} (\bibinfo {year} {2020})}\BibitemShut {NoStop}%
\bibitem [{\citenamefont {Weinfurter}(1994)}]{Weinfurter1994experimental}%
  \BibitemOpen
  \bibfield  {author} {\bibinfo {author} {\bibfnamefont {H.}~\bibnamefont {Weinfurter}},\ }\bibfield  {title} {\bibinfo {title} {Experimental bell-state analysis},\ }\href@noop {} {\bibfield  {journal} {\bibinfo  {journal} {Europhys. Lett.}\ }\textbf {\bibinfo {volume} {25}},\ \bibinfo {pages} {559} (\bibinfo {year} {1994})}\BibitemShut {NoStop}%
\bibitem [{\citenamefont {Jiang}\ \emph {et~al.}(2007)\citenamefont {Jiang}, \citenamefont {Taylor},\ and\ \citenamefont {Lukin}}]{jiang2007fast}%
  \BibitemOpen
  \bibfield  {author} {\bibinfo {author} {\bibfnamefont {L.}~\bibnamefont {Jiang}}, \bibinfo {author} {\bibfnamefont {J.~M.}\ \bibnamefont {Taylor}},\ and\ \bibinfo {author} {\bibfnamefont {M.~D.}\ \bibnamefont {Lukin}},\ }\bibfield  {title} {\bibinfo {title} {Fast and robust approach to long-distance quantum communication with atomic ensembles},\ }\href@noop {} {\bibfield  {journal} {\bibinfo  {journal} {Phys. Rev. A}\ }\textbf {\bibinfo {volume} {76}},\ \bibinfo {pages} {012301} (\bibinfo {year} {2007})}\BibitemShut {NoStop}%
\bibitem [{\citenamefont {Lu}\ \emph {et~al.}(2021)\citenamefont {Lu}, \citenamefont {Cao}, \citenamefont {Weng}, \citenamefont {Gu}, \citenamefont {Xie}, \citenamefont {Zhou}, \citenamefont {Yin},\ and\ \citenamefont {Chen}}]{Lu2021efficient}%
  \BibitemOpen
  \bibfield  {author} {\bibinfo {author} {\bibfnamefont {Y.-S.}\ \bibnamefont {Lu}}, \bibinfo {author} {\bibfnamefont {X.-Y.}\ \bibnamefont {Cao}}, \bibinfo {author} {\bibfnamefont {C.-X.}\ \bibnamefont {Weng}}, \bibinfo {author} {\bibfnamefont {J.}~\bibnamefont {Gu}}, \bibinfo {author} {\bibfnamefont {Y.-M.}\ \bibnamefont {Xie}}, \bibinfo {author} {\bibfnamefont {M.-G.}\ \bibnamefont {Zhou}}, \bibinfo {author} {\bibfnamefont {H.-L.}\ \bibnamefont {Yin}},\ and\ \bibinfo {author} {\bibfnamefont {Z.-B.}\ \bibnamefont {Chen}},\ }\bibfield  {title} {\bibinfo {title} {Efficient quantum digital signatures without symmetrization step},\ }\href@noop {} {\bibfield  {journal} {\bibinfo  {journal} {Opt. Express}\ }\textbf {\bibinfo {volume} {29}},\ \bibinfo {pages} {10162} (\bibinfo {year} {2021})}\BibitemShut {NoStop}%
\bibitem [{\citenamefont {Zeng}\ \emph {et~al.}(2022)\citenamefont {Zeng}, \citenamefont {Zhou}, \citenamefont {Wu},\ and\ \citenamefont {Ma}}]{zeng2022mode}%
  \BibitemOpen
  \bibfield  {author} {\bibinfo {author} {\bibfnamefont {P.}~\bibnamefont {Zeng}}, \bibinfo {author} {\bibfnamefont {H.}~\bibnamefont {Zhou}}, \bibinfo {author} {\bibfnamefont {W.}~\bibnamefont {Wu}},\ and\ \bibinfo {author} {\bibfnamefont {X.}~\bibnamefont {Ma}},\ }\bibfield  {title} {\bibinfo {title} {Mode-pairing quantum key distribution},\ }\href@noop {} {\bibfield  {journal} {\bibinfo  {journal} {Nat. Commun.}\ }\textbf {\bibinfo {volume} {13}},\ \bibinfo {pages} {3903} (\bibinfo {year} {2022})}\BibitemShut {NoStop}%
\bibitem [{\citenamefont {Zhou}\ \emph {et~al.}(2023)\citenamefont {Zhou}, \citenamefont {Lin}, \citenamefont {Xie}, \citenamefont {Lu}, \citenamefont {Jing}, \citenamefont {Yin},\ and\ \citenamefont {Yuan}}]{zhou2023experimental}%
  \BibitemOpen
  \bibfield  {author} {\bibinfo {author} {\bibfnamefont {L.}~\bibnamefont {Zhou}}, \bibinfo {author} {\bibfnamefont {J.}~\bibnamefont {Lin}}, \bibinfo {author} {\bibfnamefont {Y.-M.}\ \bibnamefont {Xie}}, \bibinfo {author} {\bibfnamefont {Y.-S.}\ \bibnamefont {Lu}}, \bibinfo {author} {\bibfnamefont {Y.}~\bibnamefont {Jing}}, \bibinfo {author} {\bibfnamefont {H.-L.}\ \bibnamefont {Yin}},\ and\ \bibinfo {author} {\bibfnamefont {Z.}~\bibnamefont {Yuan}},\ }\bibfield  {title} {\bibinfo {title} {Experimental quantum communication overcomes the rate-loss limit without global phase tracking},\ }\href@noop {} {\bibfield  {journal} {\bibinfo  {journal} {Phys. Rev. Lett.}\ }\textbf {\bibinfo {volume} {130}},\ \bibinfo {pages} {250801} (\bibinfo {year} {2023})}\BibitemShut {NoStop}%
\bibitem [{\citenamefont {Zhu}\ \emph {et~al.}(2023)\citenamefont {Zhu}, \citenamefont {Huang}, \citenamefont {Liu}, \citenamefont {Zeng}, \citenamefont {Zou}, \citenamefont {Dai}, \citenamefont {Tang}, \citenamefont {Li}, \citenamefont {You}, \citenamefont {Wang}, \citenamefont {Chen}, \citenamefont {Ma}, \citenamefont {Chen},\ and\ \citenamefont {Pan}}]{Zhuexp2023}%
  \BibitemOpen
  \bibfield  {author} {\bibinfo {author} {\bibfnamefont {H.-T.}\ \bibnamefont {Zhu}}, \bibinfo {author} {\bibfnamefont {Y.}~\bibnamefont {Huang}}, \bibinfo {author} {\bibfnamefont {H.}~\bibnamefont {Liu}}, \bibinfo {author} {\bibfnamefont {P.}~\bibnamefont {Zeng}}, \bibinfo {author} {\bibfnamefont {M.}~\bibnamefont {Zou}}, \bibinfo {author} {\bibfnamefont {Y.}~\bibnamefont {Dai}}, \bibinfo {author} {\bibfnamefont {S.}~\bibnamefont {Tang}}, \bibinfo {author} {\bibfnamefont {H.}~\bibnamefont {Li}}, \bibinfo {author} {\bibfnamefont {L.}~\bibnamefont {You}}, \bibinfo {author} {\bibfnamefont {Z.}~\bibnamefont {Wang}}, \bibinfo {author} {\bibfnamefont {Y.-A.}\ \bibnamefont {Chen}}, \bibinfo {author} {\bibfnamefont {X.}~\bibnamefont {Ma}}, \bibinfo {author} {\bibfnamefont {T.-Y.}\ \bibnamefont {Chen}},\ and\ \bibinfo {author} {\bibfnamefont {J.-W.}\ \bibnamefont {Pan}},\ }\bibfield  {title} {\bibinfo {title} {Experimental mode-pairing measurement-device-independent quantum key distribution without global phase
  locking},\ }\href@noop {} {\bibfield  {journal} {\bibinfo  {journal} {Phys. Rev. Lett.}\ }\textbf {\bibinfo {volume} {130}},\ \bibinfo {pages} {030801} (\bibinfo {year} {2023})}\BibitemShut {NoStop}%
\bibitem [{\citenamefont {Chou}\ \emph {et~al.}(2007)\citenamefont {Chou}, \citenamefont {Laurat}, \citenamefont {Deng}, \citenamefont {Choi}, \citenamefont {de~Riedmatten}, \citenamefont {Felinto},\ and\ \citenamefont {Kimble}}]{chou2007functional}%
  \BibitemOpen
  \bibfield  {author} {\bibinfo {author} {\bibfnamefont {C.-W.}\ \bibnamefont {Chou}}, \bibinfo {author} {\bibfnamefont {J.}~\bibnamefont {Laurat}}, \bibinfo {author} {\bibfnamefont {H.}~\bibnamefont {Deng}}, \bibinfo {author} {\bibfnamefont {K.~S.}\ \bibnamefont {Choi}}, \bibinfo {author} {\bibfnamefont {H.}~\bibnamefont {de~Riedmatten}}, \bibinfo {author} {\bibfnamefont {D.}~\bibnamefont {Felinto}},\ and\ \bibinfo {author} {\bibfnamefont {H.~J.}\ \bibnamefont {Kimble}},\ }\bibfield  {title} {\bibinfo {title} {Functional quantum nodes for entanglement distribution over scalable quantum networks},\ }\href@noop {} {\bibfield  {journal} {\bibinfo  {journal} {Science}\ }\textbf {\bibinfo {volume} {316}},\ \bibinfo {pages} {1316} (\bibinfo {year} {2007})}\BibitemShut {NoStop}%
\bibitem [{\citenamefont {Chou}\ \emph {et~al.}(2005)\citenamefont {Chou}, \citenamefont {De~Riedmatten}, \citenamefont {Felinto}, \citenamefont {Polyakov}, \citenamefont {Van~Enk},\ and\ \citenamefont {Kimble}}]{chou2005measurement}%
  \BibitemOpen
  \bibfield  {author} {\bibinfo {author} {\bibfnamefont {C.-W.}\ \bibnamefont {Chou}}, \bibinfo {author} {\bibfnamefont {H.}~\bibnamefont {De~Riedmatten}}, \bibinfo {author} {\bibfnamefont {D.}~\bibnamefont {Felinto}}, \bibinfo {author} {\bibfnamefont {S.~V.}\ \bibnamefont {Polyakov}}, \bibinfo {author} {\bibfnamefont {S.~J.}\ \bibnamefont {Van~Enk}},\ and\ \bibinfo {author} {\bibfnamefont {H.~J.}\ \bibnamefont {Kimble}},\ }\bibfield  {title} {\bibinfo {title} {Measurement-induced entanglement for excitation stored in remote atomic ensembles},\ }\href@noop {} {\bibfield  {journal} {\bibinfo  {journal} {Nature}\ }\textbf {\bibinfo {volume} {438}},\ \bibinfo {pages} {828} (\bibinfo {year} {2005})}\BibitemShut {NoStop}%
\bibitem [{\citenamefont {Slodi\ifmmode~\check{c}\else \v{c}\fi{}ka}\ \emph {et~al.}(2013)\citenamefont {Slodi\ifmmode~\check{c}\else \v{c}\fi{}ka}, \citenamefont {H\'etet}, \citenamefont {R\"ock}, \citenamefont {Schindler}, \citenamefont {Hennrich},\ and\ \citenamefont {Blatt}}]{slodic2013atom}%
  \BibitemOpen
  \bibfield  {author} {\bibinfo {author} {\bibfnamefont {L.}~\bibnamefont {Slodi\ifmmode~\check{c}\else \v{c}\fi{}ka}}, \bibinfo {author} {\bibfnamefont {G.}~\bibnamefont {H\'etet}}, \bibinfo {author} {\bibfnamefont {N.}~\bibnamefont {R\"ock}}, \bibinfo {author} {\bibfnamefont {P.}~\bibnamefont {Schindler}}, \bibinfo {author} {\bibfnamefont {M.}~\bibnamefont {Hennrich}},\ and\ \bibinfo {author} {\bibfnamefont {R.}~\bibnamefont {Blatt}},\ }\bibfield  {title} {\bibinfo {title} {Atom-atom entanglement by single-photon detection},\ }\href@noop {} {\bibfield  {journal} {\bibinfo  {journal} {Phys. Rev. Lett.}\ }\textbf {\bibinfo {volume} {110}},\ \bibinfo {pages} {083603} (\bibinfo {year} {2013})}\BibitemShut {NoStop}%
\bibitem [{\citenamefont {Humphreys}\ \emph {et~al.}(2018)\citenamefont {Humphreys}, \citenamefont {Kalb}, \citenamefont {Morits}, \citenamefont {Schouten}, \citenamefont {Vermeulen}, \citenamefont {Twitchen}, \citenamefont {Markham},\ and\ \citenamefont {Hanson}}]{humphreys2018deterministic}%
  \BibitemOpen
  \bibfield  {author} {\bibinfo {author} {\bibfnamefont {P.~C.}\ \bibnamefont {Humphreys}}, \bibinfo {author} {\bibfnamefont {N.}~\bibnamefont {Kalb}}, \bibinfo {author} {\bibfnamefont {J.~P.}\ \bibnamefont {Morits}}, \bibinfo {author} {\bibfnamefont {R.~N.}\ \bibnamefont {Schouten}}, \bibinfo {author} {\bibfnamefont {R.~F.}\ \bibnamefont {Vermeulen}}, \bibinfo {author} {\bibfnamefont {D.~J.}\ \bibnamefont {Twitchen}}, \bibinfo {author} {\bibfnamefont {M.}~\bibnamefont {Markham}},\ and\ \bibinfo {author} {\bibfnamefont {R.}~\bibnamefont {Hanson}},\ }\bibfield  {title} {\bibinfo {title} {Deterministic delivery of remote entanglement on a quantum network},\ }\href@noop {} {\bibfield  {journal} {\bibinfo  {journal} {Nature}\ }\textbf {\bibinfo {volume} {558}},\ \bibinfo {pages} {268} (\bibinfo {year} {2018})}\BibitemShut {NoStop}%
\bibitem [{\citenamefont {Delteil}\ \emph {et~al.}(2016)\citenamefont {Delteil}, \citenamefont {Sun}, \citenamefont {Gao}, \citenamefont {Togan}, \citenamefont {Faelt},\ and\ \citenamefont {Imamo{\u{g}}lu}}]{delteil2016generation}%
  \BibitemOpen
  \bibfield  {author} {\bibinfo {author} {\bibfnamefont {A.}~\bibnamefont {Delteil}}, \bibinfo {author} {\bibfnamefont {Z.}~\bibnamefont {Sun}}, \bibinfo {author} {\bibfnamefont {W.-b.}\ \bibnamefont {Gao}}, \bibinfo {author} {\bibfnamefont {E.}~\bibnamefont {Togan}}, \bibinfo {author} {\bibfnamefont {S.}~\bibnamefont {Faelt}},\ and\ \bibinfo {author} {\bibfnamefont {A.}~\bibnamefont {Imamo{\u{g}}lu}},\ }\bibfield  {title} {\bibinfo {title} {Generation of heralded entanglement between distant hole spins},\ }\href@noop {} {\bibfield  {journal} {\bibinfo  {journal} {Nat. Phys.}\ }\textbf {\bibinfo {volume} {12}},\ \bibinfo {pages} {218} (\bibinfo {year} {2016})}\BibitemShut {NoStop}%
\bibitem [{\citenamefont {Hofmann}\ \emph {et~al.}(2012)\citenamefont {Hofmann}, \citenamefont {Krug}, \citenamefont {Ortegel}, \citenamefont {Gérard}, \citenamefont {Weber}, \citenamefont {Rosenfeld},\ and\ \citenamefont {Weinfurter}}]{julian2012heralded}%
  \BibitemOpen
  \bibfield  {author} {\bibinfo {author} {\bibfnamefont {J.}~\bibnamefont {Hofmann}}, \bibinfo {author} {\bibfnamefont {M.}~\bibnamefont {Krug}}, \bibinfo {author} {\bibfnamefont {N.}~\bibnamefont {Ortegel}}, \bibinfo {author} {\bibfnamefont {L.}~\bibnamefont {Gérard}}, \bibinfo {author} {\bibfnamefont {M.}~\bibnamefont {Weber}}, \bibinfo {author} {\bibfnamefont {W.}~\bibnamefont {Rosenfeld}},\ and\ \bibinfo {author} {\bibfnamefont {H.}~\bibnamefont {Weinfurter}},\ }\bibfield  {title} {\bibinfo {title} {Heralded entanglement between widely separated atoms},\ }\href@noop {} {\bibfield  {journal} {\bibinfo  {journal} {Science}\ }\textbf {\bibinfo {volume} {337}},\ \bibinfo {pages} {72} (\bibinfo {year} {2012})}\BibitemShut {NoStop}%
\bibitem [{\citenamefont {N\"olleke}\ \emph {et~al.}(2013)\citenamefont {N\"olleke}, \citenamefont {Neuzner}, \citenamefont {Reiserer}, \citenamefont {Hahn}, \citenamefont {Rempe},\ and\ \citenamefont {Ritter}}]{nolleke2013efficient}%
  \BibitemOpen
  \bibfield  {author} {\bibinfo {author} {\bibfnamefont {C.}~\bibnamefont {N\"olleke}}, \bibinfo {author} {\bibfnamefont {A.}~\bibnamefont {Neuzner}}, \bibinfo {author} {\bibfnamefont {A.}~\bibnamefont {Reiserer}}, \bibinfo {author} {\bibfnamefont {C.}~\bibnamefont {Hahn}}, \bibinfo {author} {\bibfnamefont {G.}~\bibnamefont {Rempe}},\ and\ \bibinfo {author} {\bibfnamefont {S.}~\bibnamefont {Ritter}},\ }\bibfield  {title} {\bibinfo {title} {Efficient teleportation between remote single-atom quantum memories},\ }\href@noop {} {\bibfield  {journal} {\bibinfo  {journal} {Phys. Rev. Lett.}\ }\textbf {\bibinfo {volume} {110}},\ \bibinfo {pages} {140403} (\bibinfo {year} {2013})}\BibitemShut {NoStop}%
\bibitem [{\citenamefont {Yuan}\ \emph {et~al.}(2008)\citenamefont {Yuan}, \citenamefont {Chen}, \citenamefont {Zhao}, \citenamefont {Chen}, \citenamefont {Schmiedmayer},\ and\ \citenamefont {Pan}}]{yuan2008experimental}%
  \BibitemOpen
  \bibfield  {author} {\bibinfo {author} {\bibfnamefont {Z.-S.}\ \bibnamefont {Yuan}}, \bibinfo {author} {\bibfnamefont {Y.-A.}\ \bibnamefont {Chen}}, \bibinfo {author} {\bibfnamefont {B.}~\bibnamefont {Zhao}}, \bibinfo {author} {\bibfnamefont {S.}~\bibnamefont {Chen}}, \bibinfo {author} {\bibfnamefont {J.}~\bibnamefont {Schmiedmayer}},\ and\ \bibinfo {author} {\bibfnamefont {J.-W.}\ \bibnamefont {Pan}},\ }\bibfield  {title} {\bibinfo {title} {Experimental demonstration of a bdcz quantum repeater node},\ }\href@noop {} {\bibfield  {journal} {\bibinfo  {journal} {Nature}\ }\textbf {\bibinfo {volume} {454}},\ \bibinfo {pages} {1098} (\bibinfo {year} {2008})}\BibitemShut {NoStop}%
\bibitem [{\citenamefont {Stephenson}\ \emph {et~al.}(2020)\citenamefont {Stephenson}, \citenamefont {Nadlinger}, \citenamefont {Nichol}, \citenamefont {An}, \citenamefont {Drmota}, \citenamefont {Ballance}, \citenamefont {Thirumalai}, \citenamefont {Goodwin}, \citenamefont {Lucas},\ and\ \citenamefont {Ballance}}]{stephenson2020high}%
  \BibitemOpen
  \bibfield  {author} {\bibinfo {author} {\bibfnamefont {L.~J.}\ \bibnamefont {Stephenson}}, \bibinfo {author} {\bibfnamefont {D.~P.}\ \bibnamefont {Nadlinger}}, \bibinfo {author} {\bibfnamefont {B.~C.}\ \bibnamefont {Nichol}}, \bibinfo {author} {\bibfnamefont {S.}~\bibnamefont {An}}, \bibinfo {author} {\bibfnamefont {P.}~\bibnamefont {Drmota}}, \bibinfo {author} {\bibfnamefont {T.~G.}\ \bibnamefont {Ballance}}, \bibinfo {author} {\bibfnamefont {K.}~\bibnamefont {Thirumalai}}, \bibinfo {author} {\bibfnamefont {J.~F.}\ \bibnamefont {Goodwin}}, \bibinfo {author} {\bibfnamefont {D.~M.}\ \bibnamefont {Lucas}},\ and\ \bibinfo {author} {\bibfnamefont {C.~J.}\ \bibnamefont {Ballance}},\ }\bibfield  {title} {\bibinfo {title} {High-rate, high-fidelity entanglement of qubits across an elementary quantum network},\ }\href@noop {} {\bibfield  {journal} {\bibinfo  {journal} {Phys. Rev. Lett.}\ }\textbf {\bibinfo {volume} {124}},\ \bibinfo {pages} {110501} (\bibinfo {year} {2020})}\BibitemShut {NoStop}%
\bibitem [{\citenamefont {Hensen}\ \emph {et~al.}(2015)\citenamefont {Hensen}, \citenamefont {Bernien}, \citenamefont {Dr{\'e}au}, \citenamefont {Reiserer}, \citenamefont {Kalb}, \citenamefont {Blok}, \citenamefont {Ruitenberg}, \citenamefont {Vermeulen}, \citenamefont {Schouten}, \citenamefont {Abell{\'a}n} \emph {et~al.}}]{hensen2015loophole}%
  \BibitemOpen
  \bibfield  {author} {\bibinfo {author} {\bibfnamefont {B.}~\bibnamefont {Hensen}}, \bibinfo {author} {\bibfnamefont {H.}~\bibnamefont {Bernien}}, \bibinfo {author} {\bibfnamefont {A.~E.}\ \bibnamefont {Dr{\'e}au}}, \bibinfo {author} {\bibfnamefont {A.}~\bibnamefont {Reiserer}}, \bibinfo {author} {\bibfnamefont {N.}~\bibnamefont {Kalb}}, \bibinfo {author} {\bibfnamefont {M.~S.}\ \bibnamefont {Blok}}, \bibinfo {author} {\bibfnamefont {J.}~\bibnamefont {Ruitenberg}}, \bibinfo {author} {\bibfnamefont {R.~F.}\ \bibnamefont {Vermeulen}}, \bibinfo {author} {\bibfnamefont {R.~N.}\ \bibnamefont {Schouten}}, \bibinfo {author} {\bibfnamefont {C.}~\bibnamefont {Abell{\'a}n}}, \emph {et~al.},\ }\bibfield  {title} {\bibinfo {title} {Loophole-free bell inequality violation using electron spins separated by 1.3 kilometres},\ }\href@noop {} {\bibfield  {journal} {\bibinfo  {journal} {Nature}\ }\textbf {\bibinfo {volume} {526}},\ \bibinfo {pages} {682} (\bibinfo {year} {2015})}\BibitemShut {NoStop}%
\bibitem [{\citenamefont {Lucamarini}\ \emph {et~al.}(2018)\citenamefont {Lucamarini}, \citenamefont {Yuan}, \citenamefont {Dynes},\ and\ \citenamefont {Shields}}]{lucamarini2018overcoming}%
  \BibitemOpen
  \bibfield  {author} {\bibinfo {author} {\bibfnamefont {M.}~\bibnamefont {Lucamarini}}, \bibinfo {author} {\bibfnamefont {Z.}~\bibnamefont {Yuan}}, \bibinfo {author} {\bibfnamefont {J.}~\bibnamefont {Dynes}},\ and\ \bibinfo {author} {\bibfnamefont {A.}~\bibnamefont {Shields}},\ }\bibfield  {title} {\bibinfo {title} {Overcoming the rate--distance limit of quantum key distribution without quantum repeaters},\ }\href@noop {} {\bibfield  {journal} {\bibinfo  {journal} {Nature}\ }\textbf {\bibinfo {volume} {557}},\ \bibinfo {pages} {400} (\bibinfo {year} {2018})}\BibitemShut {NoStop}%
\bibitem [{\citenamefont {Lo}\ \emph {et~al.}(2012)\citenamefont {Lo}, \citenamefont {Curty},\ and\ \citenamefont {Qi}}]{lo2012mdi}%
  \BibitemOpen
  \bibfield  {author} {\bibinfo {author} {\bibfnamefont {H.-K.}\ \bibnamefont {Lo}}, \bibinfo {author} {\bibfnamefont {M.}~\bibnamefont {Curty}},\ and\ \bibinfo {author} {\bibfnamefont {B.}~\bibnamefont {Qi}},\ }\bibfield  {title} {\bibinfo {title} {Measurement-device-independent quantum key distribution},\ }\href@noop {} {\bibfield  {journal} {\bibinfo  {journal} {Phys. Rev. Lett.}\ }\textbf {\bibinfo {volume} {108}},\ \bibinfo {pages} {130503} (\bibinfo {year} {2012})}\BibitemShut {NoStop}%
\bibitem [{\citenamefont {Leghtas}\ \emph {et~al.}(2013)\citenamefont {Leghtas}, \citenamefont {Kirchmair}, \citenamefont {Vlastakis}, \citenamefont {Schoelkopf}, \citenamefont {Devoret},\ and\ \citenamefont {Mirrahimi}}]{leghtas2013hardware}%
  \BibitemOpen
  \bibfield  {author} {\bibinfo {author} {\bibfnamefont {Z.}~\bibnamefont {Leghtas}}, \bibinfo {author} {\bibfnamefont {G.}~\bibnamefont {Kirchmair}}, \bibinfo {author} {\bibfnamefont {B.}~\bibnamefont {Vlastakis}}, \bibinfo {author} {\bibfnamefont {R.~J.}\ \bibnamefont {Schoelkopf}}, \bibinfo {author} {\bibfnamefont {M.~H.}\ \bibnamefont {Devoret}},\ and\ \bibinfo {author} {\bibfnamefont {M.}~\bibnamefont {Mirrahimi}},\ }\bibfield  {title} {\bibinfo {title} {Hardware-efficient autonomous quantum memory protection},\ }\href@noop {} {\bibfield  {journal} {\bibinfo  {journal} {Phys. Rev. Lett.}\ }\textbf {\bibinfo {volume} {111}},\ \bibinfo {pages} {120501} (\bibinfo {year} {2013})}\BibitemShut {NoStop}%
\bibitem [{\citenamefont {Rosenblum}\ \emph {et~al.}(2018)\citenamefont {Rosenblum}, \citenamefont {Gao}, \citenamefont {Reinhold}, \citenamefont {Wang}, \citenamefont {Axline}, \citenamefont {Frunzio}, \citenamefont {Girvin}, \citenamefont {Jiang}, \citenamefont {Mirrahimi}, \citenamefont {Devoret} \emph {et~al.}}]{rosenblum2018cnot}%
  \BibitemOpen
  \bibfield  {author} {\bibinfo {author} {\bibfnamefont {S.}~\bibnamefont {Rosenblum}}, \bibinfo {author} {\bibfnamefont {Y.~Y.}\ \bibnamefont {Gao}}, \bibinfo {author} {\bibfnamefont {P.}~\bibnamefont {Reinhold}}, \bibinfo {author} {\bibfnamefont {C.}~\bibnamefont {Wang}}, \bibinfo {author} {\bibfnamefont {C.~J.}\ \bibnamefont {Axline}}, \bibinfo {author} {\bibfnamefont {L.}~\bibnamefont {Frunzio}}, \bibinfo {author} {\bibfnamefont {S.~M.}\ \bibnamefont {Girvin}}, \bibinfo {author} {\bibfnamefont {L.}~\bibnamefont {Jiang}}, \bibinfo {author} {\bibfnamefont {M.}~\bibnamefont {Mirrahimi}}, \bibinfo {author} {\bibfnamefont {M.~H.}\ \bibnamefont {Devoret}}, \emph {et~al.},\ }\bibfield  {title} {\bibinfo {title} {A cnot gate between multiphoton qubits encoded in two cavities},\ }\href@noop {} {\bibfield  {journal} {\bibinfo  {journal} {Nat. Commun.}\ }\textbf {\bibinfo {volume} {9}},\ \bibinfo {pages} {652} (\bibinfo {year} {2018})}\BibitemShut {NoStop}%
\bibitem [{\citenamefont {Kumar}\ \emph {et~al.}(2019)\citenamefont {Kumar}, \citenamefont {Lauk},\ and\ \citenamefont {Simon}}]{Kumar2019towards}%
  \BibitemOpen
  \bibfield  {author} {\bibinfo {author} {\bibfnamefont {S.}~\bibnamefont {Kumar}}, \bibinfo {author} {\bibfnamefont {N.}~\bibnamefont {Lauk}},\ and\ \bibinfo {author} {\bibfnamefont {C.}~\bibnamefont {Simon}},\ }\bibfield  {title} {\bibinfo {title} {Towards long-distance quantum networks with superconducting processors and optical links},\ }\href@noop {} {\bibfield  {journal} {\bibinfo  {journal} {Quantum Sci. Technol.}\ }\textbf {\bibinfo {volume} {4}},\ \bibinfo {pages} {045003} (\bibinfo {year} {2019})}\BibitemShut {NoStop}%
\bibitem [{\citenamefont {Goto}(2016)}]{goto2016bifurcation}%
  \BibitemOpen
  \bibfield  {author} {\bibinfo {author} {\bibfnamefont {H.}~\bibnamefont {Goto}},\ }\bibfield  {title} {\bibinfo {title} {Bifurcation-based adiabatic quantum computation with a nonlinear oscillator network},\ }\href@noop {} {\bibfield  {journal} {\bibinfo  {journal} {Sci. Rep.}\ }\textbf {\bibinfo {volume} {6}},\ \bibinfo {pages} {21686} (\bibinfo {year} {2016})}\BibitemShut {NoStop}%
\bibitem [{\citenamefont {Whaley}\ and\ \citenamefont {Milburn}(2015)}]{focus2015birgitta}%
  \BibitemOpen
  \bibfield  {author} {\bibinfo {author} {\bibfnamefont {B.}~\bibnamefont {Whaley}}\ and\ \bibinfo {author} {\bibfnamefont {G.}~\bibnamefont {Milburn}},\ }\bibfield  {title} {\bibinfo {title} {Focus on coherent control of complex quantum systems},\ }\href@noop {} {\bibfield  {journal} {\bibinfo  {journal} {New J. Phys.}\ }\textbf {\bibinfo {volume} {17}},\ \bibinfo {pages} {100202} (\bibinfo {year} {2015})}\BibitemShut {NoStop}%
\bibitem [{\citenamefont {Puri}\ \emph {et~al.}(2017)\citenamefont {Puri}, \citenamefont {Boutin},\ and\ \citenamefont {Blais}}]{puri2017engineering}%
  \BibitemOpen
  \bibfield  {author} {\bibinfo {author} {\bibfnamefont {S.}~\bibnamefont {Puri}}, \bibinfo {author} {\bibfnamefont {S.}~\bibnamefont {Boutin}},\ and\ \bibinfo {author} {\bibfnamefont {A.}~\bibnamefont {Blais}},\ }\bibfield  {title} {\bibinfo {title} {Engineering the quantum states of light in a kerr-nonlinear resonator by two-photon driving},\ }\href@noop {} {\bibfield  {journal} {\bibinfo  {journal} {npj Quantum Inf.}\ }\textbf {\bibinfo {volume} {3}},\ \bibinfo {pages} {18} (\bibinfo {year} {2017})}\BibitemShut {NoStop}%
\bibitem [{\citenamefont {Sun}\ \emph {et~al.}(2014)\citenamefont {Sun}, \citenamefont {Petrenko}, \citenamefont {Leghtas}, \citenamefont {Vlastakis}, \citenamefont {Kirchmair}, \citenamefont {Sliwa}, \citenamefont {Narla}, \citenamefont {Hatridge}, \citenamefont {Shankar}, \citenamefont {Blumoff} \emph {et~al.}}]{sun2014tracking}%
  \BibitemOpen
  \bibfield  {author} {\bibinfo {author} {\bibfnamefont {L.}~\bibnamefont {Sun}}, \bibinfo {author} {\bibfnamefont {A.}~\bibnamefont {Petrenko}}, \bibinfo {author} {\bibfnamefont {Z.}~\bibnamefont {Leghtas}}, \bibinfo {author} {\bibfnamefont {B.}~\bibnamefont {Vlastakis}}, \bibinfo {author} {\bibfnamefont {G.}~\bibnamefont {Kirchmair}}, \bibinfo {author} {\bibfnamefont {K.}~\bibnamefont {Sliwa}}, \bibinfo {author} {\bibfnamefont {A.}~\bibnamefont {Narla}}, \bibinfo {author} {\bibfnamefont {M.}~\bibnamefont {Hatridge}}, \bibinfo {author} {\bibfnamefont {S.}~\bibnamefont {Shankar}}, \bibinfo {author} {\bibfnamefont {J.}~\bibnamefont {Blumoff}}, \emph {et~al.},\ }\bibfield  {title} {\bibinfo {title} {Tracking photon jumps with repeated quantum non-demolition parity measurements},\ }\href@noop {} {\bibfield  {journal} {\bibinfo  {journal} {Nature}\ }\textbf {\bibinfo {volume} {511}},\ \bibinfo {pages} {444} (\bibinfo {year} {2014})}\BibitemShut {NoStop}%
\bibitem [{\citenamefont {Johnson}\ \emph {et~al.}(2010)\citenamefont {Johnson}, \citenamefont {Reed}, \citenamefont {Houck}, \citenamefont {Schuster}, \citenamefont {Bishop}, \citenamefont {Ginossar}, \citenamefont {Gambetta}, \citenamefont {DiCarlo}, \citenamefont {Frunzio}, \citenamefont {Girvin} \emph {et~al.}}]{johnson2010quantum}%
  \BibitemOpen
  \bibfield  {author} {\bibinfo {author} {\bibfnamefont {B.}~\bibnamefont {Johnson}}, \bibinfo {author} {\bibfnamefont {M.}~\bibnamefont {Reed}}, \bibinfo {author} {\bibfnamefont {A.~A.}\ \bibnamefont {Houck}}, \bibinfo {author} {\bibfnamefont {D.}~\bibnamefont {Schuster}}, \bibinfo {author} {\bibfnamefont {L.~S.}\ \bibnamefont {Bishop}}, \bibinfo {author} {\bibfnamefont {E.}~\bibnamefont {Ginossar}}, \bibinfo {author} {\bibfnamefont {J.}~\bibnamefont {Gambetta}}, \bibinfo {author} {\bibfnamefont {L.}~\bibnamefont {DiCarlo}}, \bibinfo {author} {\bibfnamefont {L.}~\bibnamefont {Frunzio}}, \bibinfo {author} {\bibfnamefont {S.}~\bibnamefont {Girvin}}, \emph {et~al.},\ }\bibfield  {title} {\bibinfo {title} {Quantum non-demolition detection of single microwave photons in a circuit},\ }\href@noop {} {\bibfield  {journal} {\bibinfo  {journal} {Nat. Phys.}\ }\textbf {\bibinfo {volume} {6}},\ \bibinfo {pages} {663} (\bibinfo {year} {2010})}\BibitemShut {NoStop}%
\bibitem [{\citenamefont {Welinski}\ \emph {et~al.}(2019)\citenamefont {Welinski}, \citenamefont {Woodburn}, \citenamefont {Lauk}, \citenamefont {Cone}, \citenamefont {Simon}, \citenamefont {Goldner},\ and\ \citenamefont {Thiel}}]{electron2019welinski}%
  \BibitemOpen
  \bibfield  {author} {\bibinfo {author} {\bibfnamefont {S.}~\bibnamefont {Welinski}}, \bibinfo {author} {\bibfnamefont {P.~J.~T.}\ \bibnamefont {Woodburn}}, \bibinfo {author} {\bibfnamefont {N.}~\bibnamefont {Lauk}}, \bibinfo {author} {\bibfnamefont {R.~L.}\ \bibnamefont {Cone}}, \bibinfo {author} {\bibfnamefont {C.}~\bibnamefont {Simon}}, \bibinfo {author} {\bibfnamefont {P.}~\bibnamefont {Goldner}},\ and\ \bibinfo {author} {\bibfnamefont {C.~W.}\ \bibnamefont {Thiel}},\ }\bibfield  {title} {\bibinfo {title} {Electron spin coherence in optically excited states of rare-earth ions for microwave to optical quantum transducers},\ }\href@noop {} {\bibfield  {journal} {\bibinfo  {journal} {Phys. Rev. Lett.}\ }\textbf {\bibinfo {volume} {122}},\ \bibinfo {pages} {247401} (\bibinfo {year} {2019})}\BibitemShut {NoStop}%
\bibitem [{\citenamefont {O'Brien}\ \emph {et~al.}(2014)\citenamefont {O'Brien}, \citenamefont {Lauk}, \citenamefont {Blum}, \citenamefont {Morigi},\ and\ \citenamefont {Fleischhauer}}]{interfacing2014obrien}%
  \BibitemOpen
  \bibfield  {author} {\bibinfo {author} {\bibfnamefont {C.}~\bibnamefont {O'Brien}}, \bibinfo {author} {\bibfnamefont {N.}~\bibnamefont {Lauk}}, \bibinfo {author} {\bibfnamefont {S.}~\bibnamefont {Blum}}, \bibinfo {author} {\bibfnamefont {G.}~\bibnamefont {Morigi}},\ and\ \bibinfo {author} {\bibfnamefont {M.}~\bibnamefont {Fleischhauer}},\ }\bibfield  {title} {\bibinfo {title} {Interfacing superconducting qubits and telecom photons via a rare-earth-doped crystal},\ }\href@noop {} {\bibfield  {journal} {\bibinfo  {journal} {Phys. Rev. Lett.}\ }\textbf {\bibinfo {volume} {113}},\ \bibinfo {pages} {063603} (\bibinfo {year} {2014})}\BibitemShut {NoStop}%
\bibitem [{\citenamefont {Gottesman}\ and\ \citenamefont {Chuang}(1999)}]{gottesman1999demonstrating}%
  \BibitemOpen
  \bibfield  {author} {\bibinfo {author} {\bibfnamefont {D.}~\bibnamefont {Gottesman}}\ and\ \bibinfo {author} {\bibfnamefont {I.~L.}\ \bibnamefont {Chuang}},\ }\bibfield  {title} {\bibinfo {title} {Demonstrating the viability of universal quantum computation using teleportation and single-qubit operations},\ }\href@noop {} {\bibfield  {journal} {\bibinfo  {journal} {Nature}\ }\textbf {\bibinfo {volume} {402}},\ \bibinfo {pages} {390} (\bibinfo {year} {1999})}\BibitemShut {NoStop}%
\bibitem [{\citenamefont {Deutsch}\ \emph {et~al.}(1996)\citenamefont {Deutsch}, \citenamefont {Ekert}, \citenamefont {Jozsa}, \citenamefont {Macchiavello}, \citenamefont {Popescu},\ and\ \citenamefont {Sanpera}}]{deutsch1996quantum}%
  \BibitemOpen
  \bibfield  {author} {\bibinfo {author} {\bibfnamefont {D.}~\bibnamefont {Deutsch}}, \bibinfo {author} {\bibfnamefont {A.}~\bibnamefont {Ekert}}, \bibinfo {author} {\bibfnamefont {R.}~\bibnamefont {Jozsa}}, \bibinfo {author} {\bibfnamefont {C.}~\bibnamefont {Macchiavello}}, \bibinfo {author} {\bibfnamefont {S.}~\bibnamefont {Popescu}},\ and\ \bibinfo {author} {\bibfnamefont {A.}~\bibnamefont {Sanpera}},\ }\bibfield  {title} {\bibinfo {title} {Quantum privacy amplification and the security of quantum cryptography over noisy channels},\ }\href@noop {} {\bibfield  {journal} {\bibinfo  {journal} {Phys. Rev. Lett.}\ }\textbf {\bibinfo {volume} {77}},\ \bibinfo {pages} {2818} (\bibinfo {year} {1996})}\BibitemShut {NoStop}%
\bibitem [{\citenamefont {D\"ur}\ \emph {et~al.}(1999)\citenamefont {D\"ur}, \citenamefont {Briegel}, \citenamefont {Cirac},\ and\ \citenamefont {Zoller}}]{dur1999quantum}%
  \BibitemOpen
  \bibfield  {author} {\bibinfo {author} {\bibfnamefont {W.}~\bibnamefont {D\"ur}}, \bibinfo {author} {\bibfnamefont {H.-J.}\ \bibnamefont {Briegel}}, \bibinfo {author} {\bibfnamefont {J.~I.}\ \bibnamefont {Cirac}},\ and\ \bibinfo {author} {\bibfnamefont {P.}~\bibnamefont {Zoller}},\ }\bibfield  {title} {\bibinfo {title} {Quantum repeaters based on entanglement purification},\ }\href@noop {} {\bibfield  {journal} {\bibinfo  {journal} {Phys. Rev. A}\ }\textbf {\bibinfo {volume} {59}},\ \bibinfo {pages} {169} (\bibinfo {year} {1999})}\BibitemShut {NoStop}%
\bibitem [{\citenamefont {Bennett}\ \emph {et~al.}(1996{\natexlab{a}})\citenamefont {Bennett}, \citenamefont {DiVincenzo}, \citenamefont {Smolin},\ and\ \citenamefont {Wootters}}]{bennett1996mixed}%
  \BibitemOpen
  \bibfield  {author} {\bibinfo {author} {\bibfnamefont {C.~H.}\ \bibnamefont {Bennett}}, \bibinfo {author} {\bibfnamefont {D.~P.}\ \bibnamefont {DiVincenzo}}, \bibinfo {author} {\bibfnamefont {J.~A.}\ \bibnamefont {Smolin}},\ and\ \bibinfo {author} {\bibfnamefont {W.~K.}\ \bibnamefont {Wootters}},\ }\bibfield  {title} {\bibinfo {title} {Mixed-state entanglement and quantum error correction},\ }\href@noop {} {\bibfield  {journal} {\bibinfo  {journal} {Phys. Rev. A}\ }\textbf {\bibinfo {volume} {54}},\ \bibinfo {pages} {3824} (\bibinfo {year} {1996}{\natexlab{a}})}\BibitemShut {NoStop}%
\bibitem [{\citenamefont {Bennett}\ \emph {et~al.}(1996{\natexlab{b}})\citenamefont {Bennett}, \citenamefont {Brassard}, \citenamefont {Popescu}, \citenamefont {Schumacher}, \citenamefont {Smolin},\ and\ \citenamefont {Wootters}}]{bennett1996purification}%
  \BibitemOpen
  \bibfield  {author} {\bibinfo {author} {\bibfnamefont {C.~H.}\ \bibnamefont {Bennett}}, \bibinfo {author} {\bibfnamefont {G.}~\bibnamefont {Brassard}}, \bibinfo {author} {\bibfnamefont {S.}~\bibnamefont {Popescu}}, \bibinfo {author} {\bibfnamefont {B.}~\bibnamefont {Schumacher}}, \bibinfo {author} {\bibfnamefont {J.~A.}\ \bibnamefont {Smolin}},\ and\ \bibinfo {author} {\bibfnamefont {W.~K.}\ \bibnamefont {Wootters}},\ }\bibfield  {title} {\bibinfo {title} {Purification of noisy entanglement and faithful teleportation via noisy channels},\ }\href@noop {} {\bibfield  {journal} {\bibinfo  {journal} {Phys. Rev. Lett.}\ }\textbf {\bibinfo {volume} {76}},\ \bibinfo {pages} {722} (\bibinfo {year} {1996}{\natexlab{b}})}\BibitemShut {NoStop}%
\bibitem [{\citenamefont {Briegel}\ \emph {et~al.}(1998)\citenamefont {Briegel}, \citenamefont {D\"ur}, \citenamefont {Cirac},\ and\ \citenamefont {Zoller}}]{briegel1998quantum}%
  \BibitemOpen
  \bibfield  {author} {\bibinfo {author} {\bibfnamefont {H.-J.}\ \bibnamefont {Briegel}}, \bibinfo {author} {\bibfnamefont {W.}~\bibnamefont {D\"ur}}, \bibinfo {author} {\bibfnamefont {J.~I.}\ \bibnamefont {Cirac}},\ and\ \bibinfo {author} {\bibfnamefont {P.}~\bibnamefont {Zoller}},\ }\bibfield  {title} {\bibinfo {title} {Quantum repeaters: The role of imperfect local operations in quantum communication},\ }\href@noop {} {\bibfield  {journal} {\bibinfo  {journal} {Phys. Rev. Lett.}\ }\textbf {\bibinfo {volume} {81}},\ \bibinfo {pages} {5932} (\bibinfo {year} {1998})}\BibitemShut {NoStop}%
\bibitem [{\citenamefont {Curty}\ \emph {et~al.}(2019)\citenamefont {Curty}, \citenamefont {Azuma},\ and\ \citenamefont {Lo}}]{curty2019simple}%
  \BibitemOpen
  \bibfield  {author} {\bibinfo {author} {\bibfnamefont {M.}~\bibnamefont {Curty}}, \bibinfo {author} {\bibfnamefont {K.}~\bibnamefont {Azuma}},\ and\ \bibinfo {author} {\bibfnamefont {H.-K.}\ \bibnamefont {Lo}},\ }\bibfield  {title} {\bibinfo {title} {Simple security proof of twin-field type quantum key distribution protocol},\ }\href@noop {} {\bibfield  {journal} {\bibinfo  {journal} {npj Quantum Inf.}\ }\textbf {\bibinfo {volume} {5}},\ \bibinfo {pages} {64} (\bibinfo {year} {2019})}\BibitemShut {NoStop}%
\bibitem [{\citenamefont {Abruzzo}\ \emph {et~al.}(2013)\citenamefont {Abruzzo}, \citenamefont {Bratzik}, \citenamefont {Bernardes}, \citenamefont {Kampermann}, \citenamefont {van Loock},\ and\ \citenamefont {Bru\ss{}}}]{abruzzo2013quantum}%
  \BibitemOpen
  \bibfield  {author} {\bibinfo {author} {\bibfnamefont {S.}~\bibnamefont {Abruzzo}}, \bibinfo {author} {\bibfnamefont {S.}~\bibnamefont {Bratzik}}, \bibinfo {author} {\bibfnamefont {N.~K.}\ \bibnamefont {Bernardes}}, \bibinfo {author} {\bibfnamefont {H.}~\bibnamefont {Kampermann}}, \bibinfo {author} {\bibfnamefont {P.}~\bibnamefont {van Loock}},\ and\ \bibinfo {author} {\bibfnamefont {D.}~\bibnamefont {Bru\ss{}}},\ }\bibfield  {title} {\bibinfo {title} {Quantum repeaters and quantum key distribution: Analysis of secret-key rates},\ }\href@noop {} {\bibfield  {journal} {\bibinfo  {journal} {Phys. Rev. A}\ }\textbf {\bibinfo {volume} {87}},\ \bibinfo {pages} {052315} (\bibinfo {year} {2013})}\BibitemShut {NoStop}%
\bibitem [{\citenamefont {Bratzik}\ \emph {et~al.}(2013)\citenamefont {Bratzik}, \citenamefont {Abruzzo}, \citenamefont {Kampermann},\ and\ \citenamefont {Bru\ss{}}}]{bratzik2013quantum}%
  \BibitemOpen
  \bibfield  {author} {\bibinfo {author} {\bibfnamefont {S.}~\bibnamefont {Bratzik}}, \bibinfo {author} {\bibfnamefont {S.}~\bibnamefont {Abruzzo}}, \bibinfo {author} {\bibfnamefont {H.}~\bibnamefont {Kampermann}},\ and\ \bibinfo {author} {\bibfnamefont {D.}~\bibnamefont {Bru\ss{}}},\ }\bibfield  {title} {\bibinfo {title} {Quantum repeaters and quantum key distribution: The impact of entanglement distillation on the secret key rate},\ }\href@noop {} {\bibfield  {journal} {\bibinfo  {journal} {Phys. Rev. A}\ }\textbf {\bibinfo {volume} {87}},\ \bibinfo {pages} {062335} (\bibinfo {year} {2013})}\BibitemShut {NoStop}%
\bibitem [{\citenamefont {Johansson}\ \emph {et~al.}(2012)\citenamefont {Johansson}, \citenamefont {Nation},\ and\ \citenamefont {Nori}}]{johnasson2012qutip}%
  \BibitemOpen
  \bibfield  {author} {\bibinfo {author} {\bibfnamefont {J.}~\bibnamefont {Johansson}}, \bibinfo {author} {\bibfnamefont {P.}~\bibnamefont {Nation}},\ and\ \bibinfo {author} {\bibfnamefont {F.}~\bibnamefont {Nori}},\ }\bibfield  {title} {\bibinfo {title} {Qutip: An open-source python framework for the dynamics of open quantum systems},\ }\href@noop {} {\bibfield  {journal} {\bibinfo  {journal} {Comput. Phys. Commun.}\ }\textbf {\bibinfo {volume} {183}},\ \bibinfo {pages} {1760} (\bibinfo {year} {2012})}\BibitemShut {NoStop}%
\bibitem [{\citenamefont {Machnes}\ \emph {et~al.}(2011)\citenamefont {Machnes}, \citenamefont {Sander}, \citenamefont {Glaser}, \citenamefont {de~Fouqui\`eres}, \citenamefont {Gruslys}, \citenamefont {Schirmer},\ and\ \citenamefont {Schulte-Herbr\"uggen}}]{machnes2011comparing}%
  \BibitemOpen
  \bibfield  {author} {\bibinfo {author} {\bibfnamefont {S.}~\bibnamefont {Machnes}}, \bibinfo {author} {\bibfnamefont {U.}~\bibnamefont {Sander}}, \bibinfo {author} {\bibfnamefont {S.~J.}\ \bibnamefont {Glaser}}, \bibinfo {author} {\bibfnamefont {P.}~\bibnamefont {de~Fouqui\`eres}}, \bibinfo {author} {\bibfnamefont {A.}~\bibnamefont {Gruslys}}, \bibinfo {author} {\bibfnamefont {S.}~\bibnamefont {Schirmer}},\ and\ \bibinfo {author} {\bibfnamefont {T.}~\bibnamefont {Schulte-Herbr\"uggen}},\ }\bibfield  {title} {\bibinfo {title} {Comparing, optimizing, and benchmarking quantum-control algorithms in a unifying programming framework},\ }\href@noop {} {\bibfield  {journal} {\bibinfo  {journal} {Phys. Rev. A}\ }\textbf {\bibinfo {volume} {84}},\ \bibinfo {pages} {022305} (\bibinfo {year} {2011})}\BibitemShut {NoStop}%
\bibitem [{\citenamefont {Min\'a\ifmmode~\check{r}\else \v{r}\fi{}}\ \emph {et~al.}(2008)\citenamefont {Min\'a\ifmmode~\check{r}\else \v{r}\fi{}}, \citenamefont {de~Riedmatten}, \citenamefont {Simon}, \citenamefont {Zbinden},\ and\ \citenamefont {Gisin}}]{mina2008phasenoise}%
  \BibitemOpen
  \bibfield  {author} {\bibinfo {author} {\bibfnamefont {J.~c.~v.}\ \bibnamefont {Min\'a\ifmmode~\check{r}\else \v{r}\fi{}}}, \bibinfo {author} {\bibfnamefont {H.}~\bibnamefont {de~Riedmatten}}, \bibinfo {author} {\bibfnamefont {C.}~\bibnamefont {Simon}}, \bibinfo {author} {\bibfnamefont {H.}~\bibnamefont {Zbinden}},\ and\ \bibinfo {author} {\bibfnamefont {N.}~\bibnamefont {Gisin}},\ }\bibfield  {title} {\bibinfo {title} {Phase-noise measurements in long-fiber interferometers for quantum-repeater applications},\ }\href@noop {} {\bibfield  {journal} {\bibinfo  {journal} {Phys. Rev. A}\ }\textbf {\bibinfo {volume} {77}},\ \bibinfo {pages} {052325} (\bibinfo {year} {2008})}\BibitemShut {NoStop}%
\bibitem [{\citenamefont {Sinclair}\ \emph {et~al.}(2014)\citenamefont {Sinclair}, \citenamefont {Saglamyurek}, \citenamefont {Mallahzadeh}, \citenamefont {Slater}, \citenamefont {George}, \citenamefont {Ricken}, \citenamefont {Hedges}, \citenamefont {Oblak}, \citenamefont {Simon}, \citenamefont {Sohler},\ and\ \citenamefont {Tittel}}]{sinclair2014spectral}%
  \BibitemOpen
  \bibfield  {author} {\bibinfo {author} {\bibfnamefont {N.}~\bibnamefont {Sinclair}}, \bibinfo {author} {\bibfnamefont {E.}~\bibnamefont {Saglamyurek}}, \bibinfo {author} {\bibfnamefont {H.}~\bibnamefont {Mallahzadeh}}, \bibinfo {author} {\bibfnamefont {J.~A.}\ \bibnamefont {Slater}}, \bibinfo {author} {\bibfnamefont {M.}~\bibnamefont {George}}, \bibinfo {author} {\bibfnamefont {R.}~\bibnamefont {Ricken}}, \bibinfo {author} {\bibfnamefont {M.~P.}\ \bibnamefont {Hedges}}, \bibinfo {author} {\bibfnamefont {D.}~\bibnamefont {Oblak}}, \bibinfo {author} {\bibfnamefont {C.}~\bibnamefont {Simon}}, \bibinfo {author} {\bibfnamefont {W.}~\bibnamefont {Sohler}},\ and\ \bibinfo {author} {\bibfnamefont {W.}~\bibnamefont {Tittel}},\ }\bibfield  {title} {\bibinfo {title} {Spectral multiplexing for scalable quantum photonics using an atomic frequency comb quantum memory and feed-forward control},\ }\href@noop {} {\bibfield  {journal} {\bibinfo  {journal} {Phys. Rev. Lett.}\ }\textbf {\bibinfo {volume} {113}},\ \bibinfo
  {pages} {053603} (\bibinfo {year} {2014})}\BibitemShut {NoStop}%
\bibitem [{\citenamefont {Grimau~Puigibert}\ \emph {et~al.}(2017)\citenamefont {Grimau~Puigibert}, \citenamefont {Aguilar}, \citenamefont {Zhou}, \citenamefont {Marsili}, \citenamefont {Shaw}, \citenamefont {Verma}, \citenamefont {Nam}, \citenamefont {Oblak},\ and\ \citenamefont {Tittel}}]{grimau2017heralded}%
  \BibitemOpen
  \bibfield  {author} {\bibinfo {author} {\bibfnamefont {M.}~\bibnamefont {Grimau~Puigibert}}, \bibinfo {author} {\bibfnamefont {G.~H.}\ \bibnamefont {Aguilar}}, \bibinfo {author} {\bibfnamefont {Q.}~\bibnamefont {Zhou}}, \bibinfo {author} {\bibfnamefont {F.}~\bibnamefont {Marsili}}, \bibinfo {author} {\bibfnamefont {M.~D.}\ \bibnamefont {Shaw}}, \bibinfo {author} {\bibfnamefont {V.~B.}\ \bibnamefont {Verma}}, \bibinfo {author} {\bibfnamefont {S.~W.}\ \bibnamefont {Nam}}, \bibinfo {author} {\bibfnamefont {D.}~\bibnamefont {Oblak}},\ and\ \bibinfo {author} {\bibfnamefont {W.}~\bibnamefont {Tittel}},\ }\bibfield  {title} {\bibinfo {title} {Heralded single photons based on spectral multiplexing and feed-forward control},\ }\href@noop {} {\bibfield  {journal} {\bibinfo  {journal} {Phys. Rev. Lett.}\ }\textbf {\bibinfo {volume} {119}},\ \bibinfo {pages} {083601} (\bibinfo {year} {2017})}\BibitemShut {NoStop}%
\bibitem [{\citenamefont {Bruzewicz}\ \emph {et~al.}(2019)\citenamefont {Bruzewicz}, \citenamefont {Chiaverini}, \citenamefont {McConnell},\ and\ \citenamefont {Sage}}]{bruzewicz2019trapped}%
  \BibitemOpen
  \bibfield  {author} {\bibinfo {author} {\bibfnamefont {C.~D.}\ \bibnamefont {Bruzewicz}}, \bibinfo {author} {\bibfnamefont {J.}~\bibnamefont {Chiaverini}}, \bibinfo {author} {\bibfnamefont {R.}~\bibnamefont {McConnell}},\ and\ \bibinfo {author} {\bibfnamefont {J.~M.}\ \bibnamefont {Sage}},\ }\bibfield  {title} {\bibinfo {title} {Trapped-ion quantum computing: Progress and challenges},\ }\href@noop {} {\bibfield  {journal} {\bibinfo  {journal} {Appl. Phys. Rev.}\ }\textbf {\bibinfo {volume} {6}},\ \bibinfo {pages} {021314} (\bibinfo {year} {2019})}\BibitemShut {NoStop}%
\bibitem [{\citenamefont {Brown}\ \emph {et~al.}(2011)\citenamefont {Brown}, \citenamefont {Wilson}, \citenamefont {Colombe}, \citenamefont {Ospelkaus}, \citenamefont {Meier}, \citenamefont {Knill}, \citenamefont {Leibfried},\ and\ \citenamefont {Wineland}}]{brown2021single}%
  \BibitemOpen
  \bibfield  {author} {\bibinfo {author} {\bibfnamefont {K.~R.}\ \bibnamefont {Brown}}, \bibinfo {author} {\bibfnamefont {A.~C.}\ \bibnamefont {Wilson}}, \bibinfo {author} {\bibfnamefont {Y.}~\bibnamefont {Colombe}}, \bibinfo {author} {\bibfnamefont {C.}~\bibnamefont {Ospelkaus}}, \bibinfo {author} {\bibfnamefont {A.~M.}\ \bibnamefont {Meier}}, \bibinfo {author} {\bibfnamefont {E.}~\bibnamefont {Knill}}, \bibinfo {author} {\bibfnamefont {D.}~\bibnamefont {Leibfried}},\ and\ \bibinfo {author} {\bibfnamefont {D.~J.}\ \bibnamefont {Wineland}},\ }\bibfield  {title} {\bibinfo {title} {Single-qubit-gate error below ${\mathbf{10}}^{\ensuremath{-}\mathbf{4}}$ in a trapped ion},\ }\href@noop {} {\bibfield  {journal} {\bibinfo  {journal} {Phys. Rev. A}\ }\textbf {\bibinfo {volume} {84}},\ \bibinfo {pages} {030303} (\bibinfo {year} {2011})}\BibitemShut {NoStop}%
\bibitem [{\citenamefont {Benhelm}\ \emph {et~al.}(2008)\citenamefont {Benhelm}, \citenamefont {Kirchmair}, \citenamefont {Roos},\ and\ \citenamefont {Blatt}}]{benhelm2008towards}%
  \BibitemOpen
  \bibfield  {author} {\bibinfo {author} {\bibfnamefont {J.}~\bibnamefont {Benhelm}}, \bibinfo {author} {\bibfnamefont {G.}~\bibnamefont {Kirchmair}}, \bibinfo {author} {\bibfnamefont {C.~F.}\ \bibnamefont {Roos}},\ and\ \bibinfo {author} {\bibfnamefont {R.}~\bibnamefont {Blatt}},\ }\bibfield  {title} {\bibinfo {title} {Towards fault-tolerant quantum computing with trapped ions},\ }\href@noop {} {\bibfield  {journal} {\bibinfo  {journal} {Nat. Phys.}\ }\textbf {\bibinfo {volume} {4}},\ \bibinfo {pages} {463} (\bibinfo {year} {2008})}\BibitemShut {NoStop}%
\bibitem [{\citenamefont {Myerson}\ \emph {et~al.}(2008)\citenamefont {Myerson}, \citenamefont {Szwer}, \citenamefont {Webster}, \citenamefont {Allcock}, \citenamefont {Curtis}, \citenamefont {Imreh}, \citenamefont {Sherman}, \citenamefont {Stacey}, \citenamefont {Steane},\ and\ \citenamefont {Lucas}}]{myerson2008high}%
  \BibitemOpen
  \bibfield  {author} {\bibinfo {author} {\bibfnamefont {A.~H.}\ \bibnamefont {Myerson}}, \bibinfo {author} {\bibfnamefont {D.~J.}\ \bibnamefont {Szwer}}, \bibinfo {author} {\bibfnamefont {S.~C.}\ \bibnamefont {Webster}}, \bibinfo {author} {\bibfnamefont {D.~T.~C.}\ \bibnamefont {Allcock}}, \bibinfo {author} {\bibfnamefont {M.~J.}\ \bibnamefont {Curtis}}, \bibinfo {author} {\bibfnamefont {G.}~\bibnamefont {Imreh}}, \bibinfo {author} {\bibfnamefont {J.~A.}\ \bibnamefont {Sherman}}, \bibinfo {author} {\bibfnamefont {D.~N.}\ \bibnamefont {Stacey}}, \bibinfo {author} {\bibfnamefont {A.~M.}\ \bibnamefont {Steane}},\ and\ \bibinfo {author} {\bibfnamefont {D.~M.}\ \bibnamefont {Lucas}},\ }\bibfield  {title} {\bibinfo {title} {High-fidelity readout of trapped-ion qubits},\ }\href@noop {} {\bibfield  {journal} {\bibinfo  {journal} {Phys. Rev. Lett.}\ }\textbf {\bibinfo {volume} {100}},\ \bibinfo {pages} {200502} (\bibinfo {year} {2008})}\BibitemShut {NoStop}%
\bibitem [{\citenamefont {Yoshida}\ and\ \citenamefont {Horikiri}(2024)}]{yoshida2024multiplexed}%
  \BibitemOpen
  \bibfield  {author} {\bibinfo {author} {\bibfnamefont {D.}~\bibnamefont {Yoshida}}\ and\ \bibinfo {author} {\bibfnamefont {T.}~\bibnamefont {Horikiri}},\ }\bibfield  {title} {\bibinfo {title} {Multiplexed quantum repeaters based on single-photon interference with mild stabilization},\ }\href@noop {} {\bibfield  {journal} {\bibinfo  {journal} {Commun. Phys.}\ }\textbf {\bibinfo {volume} {7}},\ \bibinfo {pages} {367} (\bibinfo {year} {2024})}\BibitemShut {NoStop}%
\bibitem [{\citenamefont {Simon}\ and\ \citenamefont {Irvine}(2003)}]{simon2003robust}%
  \BibitemOpen
  \bibfield  {author} {\bibinfo {author} {\bibfnamefont {C.}~\bibnamefont {Simon}}\ and\ \bibinfo {author} {\bibfnamefont {W.~T.~M.}\ \bibnamefont {Irvine}},\ }\bibfield  {title} {\bibinfo {title} {Robust long-distance entanglement and a loophole-free bell test with ions and photons},\ }\href@noop {} {\bibfield  {journal} {\bibinfo  {journal} {Phys. Rev. Lett.}\ }\textbf {\bibinfo {volume} {91}},\ \bibinfo {pages} {110405} (\bibinfo {year} {2003})}\BibitemShut {NoStop}%
\end{thebibliography}

\end{document}